\documentclass[a4paper,fleqn,usenatbib]{mnras}
\usepackage{color}
\usepackage{float}
\usepackage{graphicx}
\usepackage{epstopdf}
\usepackage{subcaption}
\usepackage{newtxtext,newtxmath}
\usepackage{lastpage}

\usepackage[T1]{fontenc}
\usepackage{ae,aecompl}

\newcommand{\fsed}{f_{\textrm{sed}}}

\newcommand{\alu}{$Al$_2$O$_3}
\newcommand{\mso}{$MgSiO$_3}
\newcommand{\msotwo}{$Mg$_2$SiO$_4}

\title[Radiative 3D EddySed Clouds]{Overcast on Osiris: 3D radiative--hydrodynamical simulations of a cloudy hot Jupiter using the parameterised, phase--equilibrium cloud formation code $EddySed$}

\author[S. Lines et al.]{
S. Lines,$^{1}\thanks{E-mail: s.lines@exeter.ac.uk}$
N. J. Mayne,$^{1}$
J. Manners,$^{1,3}$
I. A. Boutle,$^{1,3}$\newauthor
B. Drummond,$^{1}$
T. Mikal-Evans,$^{4}$
K. Kohary,$^{1,2}$
and D. K. Sing$^{5}$     
\\ \\
$^{1}$Physics and Astronomy, College of Engineering, Mathematics and Physical Sciences, University of Exeter, EX4 4QL\\
$^{2}$Computer Science, College of Engineering, Mathematics and Physical Sciences, University of Exeter, EX4 4QF\\
$^{3}$Met Office, FitzRoy Road, Exeter, Devon EX1 3PB, UK\\
$^{4}$Kavli Institute for Astrophysics and Space Research, Massachusetts Institute of Technology, 77 Massachusetts Avenue, 37-241, Cambridge, MA 02139, USA\\
$^{5}$Department of Earth and Planetary Sciences, Johns Hopkins University, Baltimore, MD, USA
}

\date{Accepted 26th June 2019}

\pubyear{2018}

\begin{document}
\label{firstpage}
\pagerange{\pageref{firstpage}--\pageref{lastpage}}
\maketitle

\begin{abstract}
We present results from 3D radiative--hydrodynamical simulations of HD~209458b with a fully coupled treatment of clouds using the EddySed code, critically, including cloud radiative feedback via absorption and scattering. We demonstrate that the thermal and optical structure of the simulated atmosphere is markedly different, for the majority of our simulations, when including cloud radiative effects, suggesting this important mechanism can not be neglected. Additionally, we further demonstrate that the cloud structure is sensitive to not only the cloud sedimentation efficiency (termed $\fsed$ in EddySed), but also the temperature--pressure profile of the deeper atmosphere. We briefly discuss the large difference between the resolved cloud structures of this work, adopting a phase--equilibrium and parameterised cloud model, and our previous work incorporating a cloud microphysical model, although a fairer comparison where, for example, the same list of constituent condensates is included in both treatments, is reserved for a future work. Our results underline the importance of further study into the potential condensate size distributions and vertical structures, as both strongly influence the radiative impact of clouds on the atmosphere. Finally, we present synthetic observations from our simulations reporting an improved match, over our previous cloud--free simulations, to the observed transmission, HST WFC3 emission and 4.5\,$\mu$m {\emph{Spitzer}} phase curve of HD~209458b. Additionally, we find all our cloudy simulations have an apparent albedo consistent with observations.
\end{abstract}

\begin{keywords}
methods: numerical -- hydrodynamics -- radiative transfer -- scattering -- Planets and satellites: atmospheres -- Planets and satellites: gaseous planets
\vspace{+10pt}
\end{keywords}


\section{Introduction}
\label{sec:intro}
Evidence for clouds in the atmospheres of exoplanets, particularly hot Jupiters, comes from a variety of sources spanning `super--Raleigh' scattering in the optical and near--UV wavelengths of transmission spectra, attenuated or fully masked molecular and atomic spectral features in transmission such as a weakened presence of water vapour, sodium and potassium \citep{etangs08,deming13,nikolov15,sing16,iyer16,kirk17}, and variability or offsets in their phase curves \citep{demory13,armstrong16}. However, there remains significant uncertainty surrounding the formation mechanisms, chemical compositions, particle sizes, vertical structures and transport of clouds in exoplanet atmospheres. Although often modelled as passive tracers, cloud particles or droplets have been shown via theoretical studies of both Earth's climate \citep{ramanathan89,hartmann92} and exoplanets \citep{parmentier16,lee16,lines18a,lines18b,powell18,roman18} to interact strongly with both the outgoing thermal planetary emission and stellar irradiation, feeding back into their host atmospheres, changing their thermochemical structures and subsequently imprinting their signature on observables.

Recently there has been significant progress in the application of cloud models to non--terrestrial exoplanet atmospheres, with simulations encompassing both 1D \citep[e.g.][]{helling06,benneke12,helling13,lee14,marley15,morley15,barstow17,pinhas17,charnay18,gao18a,gao18b,moran18,ohno18,ormel18,powell18} and 3D \citep[e.g.][]{parmentier13,lee16,oreshenko16,parmentier16,boutle17,lines18a,lines18b,roman18} geometries. Such work has improved our understanding of the cloud formation pathways, their influence on observed spectra, as well as the potential physical properties of cloud particles and droplets themselves, such as their chemical compositions, radii, radiative interactions, and sedimentation efficiencies. Within these works there exists a significant variation in the level of model complexity, ranging from 1D simulations with highly parameterised or prescribed and radiatively inactive cloud, to 3D simulations with comprehensive microphysics and cloud radiative feedback. Just as some approaches may over--simplify and even ignore mandatory physics, others could include extraneous physical processes, at large computational expense, and which do not impact the observations. However, in order to determine the minimum level of physical completeness required to interpret the current observations, and provide robust predictions for future observations, such a range of approaches is required. 

The most comprehensive and self--consistent simulations of hot Jupiter exoplanetary clouds to-date were performed by \citet{lines18a}, where, following the pioneering work of \citet{lee16}, a kinetic, non--equilibrium, microphysical cloud formation model \citep[see][]{helling13} was coupled to a 3D radiative--hydrodynamical model, in our case the Unified Model (UM), a verified 3D General Circulation Model (GCM) from the UK Met Office. This coupled model considers homogeneous nucleation of TiO$_2$ seed particles from the gas phase, mixed--composition surface growth (condensation), evaporation, gravitational settling (precipitation), cloud particle advection as well as gas and cloud absorption and, in an improvement over the previous approaches, scattering. \citet{lines18a}, for a simulated hot Jupiter atmosphere, showed that an abundance of TiO$_2$ cloud condensation nuclei resulted in a large abundance of sub--micron, mixed--composition cloud particles which, in turn, due to their small radii were able to suspend themselves against precipitation. Due to their size (sub-micron) and composition (silicate--based), these particles contributed a strong scattering opacity at the lowest simulated pressures, reducing the insolation to the layers directly below and cooling the upper atmosphere, thereby supporting further cloud formation. The combined effect of a latitudinal temperature variation, set by the preferential energy deposition at the sub--stellar (equatorial) point, and a meridional flow of cloud particles (and metal rich gas) out of the jet and to higher latitudes, resulted in a latitudinal banding of the cloud particles. \citet{lines18a} also found that their simulations returned an eastwards shift in the peak of the long--wave emission, or hot--spot, qualitatively in line with observations \citep{knutson07}, alongside supporting turbulent cloud structures stimulating variability in the thermal emission. However, since the atmospheric temperatures for pressures below 1\,bar did not exceed the condensation temperature of the mixed--composition silicate cloud, cloud particles were able to form and persist across all longitudes, meaning the simulations did not return a westward shift in the peak short--wave flux, as found in observations \citep[such as the Kepler-7b study of][]{demory13}. \citet{lines18b} analysed the results of \citet{lines18a} further by implementing a self--consistent 3D transmission model, finding that the high--opacity suspended silicate particles led to a transmission spectrum much flatter than suggested by observational data from \citet{sing16}.

Complex 3D microphysical simulations such as those performed in \citet{lines18a}, which include time--dependent condensed--phase chemistry and an explicit handling of cloud particle precipitation, are limited in their ability to capture the required timescales due to the computational cost of the simulations. Despite this difficulty, 3D simulations are required to capture the inherently three--dimensional structure and movement of cloud \citep{lee16,lines18a,roman18}, as well as strong inhomogeneities in atmospheric properties (such as the asymmetric stellar heating in tidally locked planets). Broad parameterisations from simplistic cloud models can hide the underlying cloud formation and evolution physics. However, if sensible comparisons are made against more complete microphysics schemes and observations, such models may be used to help constrain the cloud properties of the atmosphere, using considerably less computational resource than required for solving the phase--disequilibrium calculations within a microphysics approach. Limitations may apply however; in a comparison study of cloud schemes in 1D brown--dwarf atmospheres, \citet{helling08c} found that despite the overall distribution of cloud being well matched between approaches of differing complexity, the opacity driving properties (such as the particle sizes and composition) can vary substantially.

Here we couple the \citet{ackerman01} or `EddySed' cloud formation code to the same 3D model used in \citet{lines18b,lines18a}, and again include a full treatment of cloud radiative feedback. The EddySed cloud formation code is a respected tool in the brown--dwarf and exoplanet community and has been used widely in both 1D forward and retrieval models of sub--stellar objects \citep[e.g.][]{fortney06,cushing10,morley15,marley15,rajan17}. Given the wide range of unknowns pertaining to cloud physics, it simplifies the complex microphysical processes by implementing assumptions, such as phase--equilibrium, homogeneous condensation, and parameterisations of the particles sizes and vertical transport (diffusive up--mixing and precipitation). Such approximations can be used to imitate conditions provided by more complex microphysics models, such as in \citet{gao18a}, using the CARMA model \citep{toon79}, where the sedimentation efficiency factor, $\fsed$, is fit to a combination of detailed microphysics conditions such as condensate surface energies, seed radii and condensation nuclei flux. In this work, we investigate the impact of 3D radiatively active cloud, prescribed by the \citet{ackerman01} EddySed model, on the atmosphere of a hot Jupiter, HD~209458b. By performing our simulations using the same radiative--hydrodynamic conditions as \citet{lines18a}, we can begin to understand the differences that arise by neglecting some of the more complex cloud processes such as seed nucleation, phase--disequilibrium and true sedimentation (advection) of cloud particles.

The following study is structured as follows: in Section \ref{sec:theory} we describe the model components, initial conditions and methodology. In Section \ref{sec:resultsdis} we discuss the main results of our simulations revealing that for most cases the impact of the cloud radiative feedback can not be neglected (Section \ref{subsec:rad_clouds}), that the resulting cloud structures are dependent on the assumed sedimentation parameter and temperature--pressure profile of the deep atmosphere (Section \ref{subsec:cloud_prop}), and compare synthetic with real observations (Section \ref{subsec:synth_obs}). We then detail our conclusions in Section \ref{sec:conclusions}.


\section{Numerical Methods}
\label{sec:theory}

\begin {table*}
\begin{center}
\begin{tabular}{ l|c|c }
{\bf{Grid $\&$ Timestepping}} & {\bf{Hot Deep Interior (HDI)}} & {\bf{Standard Deep Interior (SDI)}}\\
\hline
Horizontal resolution ($\lambda$--Longitude and $\phi$--Latitude cells) & $\lambda$ = 144, $\phi$ = 90 & $\leftarrow$ \\ 
Vertical resolution (levels) & 66 & $\leftarrow$\\ 
Hydrodynamical timestep, $\tau_{\textrm{hydro}}$ (s) & 30 & $\leftarrow$ \\ 
Cloud chemistry timestep, $\tau_{\textrm{cloud}}$ (s) & 300 & $\leftarrow$\\
Radiative timestep, $\tau_{\textrm{rad}}$ (s) & 300 & $\leftarrow$\\
Total simulation time, t$_{\textrm{total}}$ (days) & 500 & $\leftarrow$ \\ \\
{\bf{Radiative Transfer Properties}}& & \\
\hline
UM wavelength bins (low--resolution) & 32 & $\leftarrow$\\ 
UM wavelength bins (high--resolution) & 500 & $\leftarrow$\\ 
Wavelength minimum, $\lambda_{\textrm{min}}$ (${\mu}$m) & 0.2 & $\leftarrow$\\
Wavelength maximum, $\lambda_{\textrm{max}}$ (${\mu}$m) & 300 & $\leftarrow$\\
EddySed wavelength bins & 192 & $\leftarrow$\\ 
EddySed particle radius bins & 40 & $\leftarrow$\\
EddySed particle radius upper ($\mu$m) & 2800 & $\leftarrow$\\
EddySed particle radius lower ($\mu$m) & 0.1 & $\leftarrow$\\ \\
{\bf{Hydrodynamical Damping}} & & \\
\hline
Damping coefficient & 0.15 & $\leftarrow$ \\
Damping geometry & Linear & $\leftarrow$ \\
Diffusion coefficient & 0.158 & $\leftarrow$\\ \\
{\bf{Planetary Constants}} & & \\
\hline
Initial inner boundary pressure (mbar) & 2.0 $\times$ 10$^5$ & $\leftarrow$ \\
Atmosphere upper boundary height (m) & 1.0 $\times$ 10$^7$ & 9.0 $\times$ 10$^6$ \\
Intrinsic temperature (K) & 100 & $\leftarrow$\\
Ideal gas constant, $R$ (Jkg$^{-1}$K$^{-1}$) & 3556.8 & $\leftarrow$ \\
Specific heat capacity, $c_{\textrm{p}}$ (Jkg$^{-1}$K$^{-1}$) & 1.3 $\times$ 10$^4$ & $\leftarrow$ \\
Planetary radius, $R_{\textrm{p}}$ (m) & 9.00 $\times$ 10$^7$ & $\leftarrow$ \\
Planetary rotation rate, $\Omega$ (s$^{-1}$) & 2.06 $\times$ 10$^{-5}$ & $\leftarrow$ \\
Surface gravity, $g_{\textrm{p}}$ (ms$^{-2}$) & 10.79 & $\leftarrow$ \\
Semi-major axis, $a_{\textrm{p}}$ (au) & 4.75 $\times$ 10$^{-2}$ & $\leftarrow$ \\ \\
{\bf{Cloud Model Parameters}} & & \\
\hline
EddySed effective temperature, $T_{\textrm{eff}}$ (K) &  1130 & $\leftarrow$ \\
Sedimentation parameter, $\fsed$ & 0.1, 1.0 & $\leftarrow$ \\
Minimum Eddy Diffusion coefficient, ($K_{\textrm{zz}}^{\textrm{min}}$, m$^2$/s) & 10 & $\leftarrow$\\
Minimum ratio of turbulent mixing length\\ to atmospheric scale height, $\Lambda$ & 0.1 & $\leftarrow$\\
Supersaturation remaining after condensation & 0 & $\leftarrow$\\
Geometric standard deviation in lognormal\\ size distribution of condensates & 2 & $\leftarrow$\\
Initial cloud opacity scaling factor & 100.0 & $\leftarrow$ \\
Cloud opacity ramping time, t$_{\textrm{ramp}}$ (days) & 100 & $\leftarrow$ \\
Included condensates & Al$_2$O$_3$, Fe, Na$_2$S, NH$_3$, KCl & $\leftarrow$ \\
 & MnS, ZnS, Cr, MgSiO$_3$ $\&$ Mg$_2$SiO$_4$ & \\
\end{tabular}
\vspace{+10pt}
\caption{Selected model parameters from both our standard and hot deep interior HD~209458b atmospheres, covering grid setup, timestepping, radiative transfer properties, run--lengths, hydrodynamical damping coefficients and planet constants. See \citet{lines18a} and references therein for more information.}
\label{tab:params}
\end{center}
\end{table*}

In this section we first introduce the main components of our model, then the practical aspects of performing the simulations (e.g., initial conditions and model parameters). Table \ref{tab:params} presents the model parameters for our simulations such as the spatial (vertical and horizontal grid) and temporal (hydrodynamical, radiative and cloud chemistry time--stepping) resolutions, radiative transfer properties, hydrodynamical damping coefficients, the \citet{ackerman01} cloud scheme parameters and a complete list of planetary constants. Apart from the cloud model settings that apply only to this work, all simulation parameters are chosen to match \citet{lines18a} unless where stated in the following text.

\subsection{Model}
\label{subsec:model}

Our simulations of HD~209458b were performed using a well tested GCM, the Met Office UM. This model has been used in previous works to simulate the atmospheres of hot Jupiters \citep{mayne14,mayne17,amundsen16,lines18a,drummond18b,drummond18c}, small Neptunes \citep{drummond18a,mayne19} and smaller terrestrial exoplanets \citep{mayne13,boutle17,lewis18}. The UM is set up to solve the full, deep--atmosphere and non--hydrostatic Navier--Stokes equations \citep{wood14,mayne14,mayne17}, and we adopt similar initial conditions and simulations parameters to \citet{lines18a,lines18b}, except where explained in this work.

We use the open source, two--stream solver `Suite Of Community RAdiative Transfer codes based on \citet{edwards96a}' (SOCRATES\footnote{https://code.metoffice.gov.uk/trac/socrates}) for the calculation of radiative heating rates, implemented in the configuration described in \citet{amundsen16}. The effects of Rayleigh scattering from a H$_2$/He dominated atmosphere are included, and the practical improved flux method \citep{zdunkowski80,zdunkowski85} is used to treat both the scattering of  stellar and thermal fluxes. The correlated--$k$ method is used for gas absorption with absorption line data for H$_2$O, CO, CH$_4$, NH$_3$, Li, Na, K, Rb, Cs and H$_2$-H$_2$ and H$_2$-He collision induced absorption (CIA) data are taken from ExoMol, and where necessary, HITRAN and HITEMP. A complete account of the line list and partition function sources can be found in \citet{amundsen14}. Finally, the equivalent extinction method \citep[see][]{edwards96b,amundsen17} is used for the treatment of overlapping gas phase absorption, as opacities of the gas mixture are calculated by mixing individual opacities at runtime. Gas--phase chemical equilibrium abundances are obtained using both the analytical \cite{burrows99} chemistry scheme and a simple temperature--dependent parameterisation for the alkali abundances, implemented as per \cite{amundsen16} for computational efficiency.

We couple to the UM the parameterised and phase--equilibrium \citet{ackerman01} cloud formation scheme, EddySed\footnote{EddySed was obtained via private communication with Mark Marley and Tiffany Kataria.}, which solves for condensation cloud properties (e.g. cloud mixing ratio, particle sizes, radiative coefficients) for a given atmospheric column by enforcing a balance between the upwards vertical transport of condensate and vapour against the downwards sedimentation (precipitation) of condensed material using the following equation:

\begin{equation}
-K_{\textrm{zz}}\frac{\partial q_{\textrm{t}}}{\partial z} = f_{\textrm{sed}}w_{*}q_{\textrm{c}}.
\label{eq:eddy}
\end{equation}
Here, $K_{\textrm{zz}}$ is the eddy diffusion coefficient, $q_{\textrm{c}}$ is the condensate mixing ratio, $q_{\textrm{t}}$ = $q_{\textrm{c}}$ + $q_{\textrm{v}}$ (the total mixing ratio of condensate and vapour, $q_{\textrm{v}}$), $z$ is the atmospheric vertical height, $\fsed$ is the free sedimentation efficiency factor (defined as the ratio of the mass--weighted cloud particle or droplet sedimentation velocity to $w_{*}$), $w_{*}$ is the convective velocity scale defined from mixing length theory as $w_{*}$ = $K_{\textrm{zz}}/L$, where the turbulent mixing length $L$ is defined as:

\begin{equation}
L = H\,\textrm{max}\left(\Lambda,\frac{\Gamma}{\Gamma_{\textrm{adiab}}}\right).
\label{eq:turb}
\end{equation}
Here, $\Lambda$ is the minimum ratio of turbulent mixing length to atmospheric scale height (given in Table \ref{tab:params}), $\Gamma$ and $\Gamma_{\textrm{adiab}}$ are the local and dry adiabatic lapse rates respectively, and $H$ is the local atmospheric scale height $H$ = $RT/g$, where T is the layer temperature, and g is the acceleration due to gravity in that layer.

To approximate the vertical mixing, EddySed uses the \citet{gierasch85} definition of the eddy diffusion coefficient:

\begin{equation}
K_{\textrm{zz}} = \textrm{max}\left(K_{\textrm{zz}}^{\textrm{min}},\frac{H}{3} \left( \frac{L}{H} \right)^{4/3} \left( \frac{RF}{\mu \rho_{a} c_{p}} \right)^{1/3}\right),
\label{kzz}
\end{equation}
where $K_{\textrm{zz}}^{\textrm{min}}$ is the minimum enforced value (to represent circulation--driven vertical advection in radiative regions, where mixing--length theory can under--predict vertical mixing) of the eddy diffusion coefficient (Table \ref{tab:params} details the settings of our simulations including this value), $R$ is the universal gas constant, $F$ is the convective heat flux where $F$ = $\sigma T_{\textrm{eff}}^4$ and T$_{\textrm{eff}}$ is the planetary effective temperature, $\mu$ is the atmospheric molecular weight (2.2 gmol$^{-1}$), $\rho_{a}$ is the atmospheric density and $c_{p}$ is the specific heat of the atmosphere at constant pressure (1.3 $\times$ 10$^{-4}$ JK$^{-1}$). This definition of $K_{\textrm{zz}}$ is a parameterisation of the convective processes in the atmosphere, which are approximated through the temperature--pressure profile. We therefore neglect the large scale flows which are not driven by convection, and caution that the vertical mixing may be stronger for the true solution. In this work, we consciously choose to adopt the \cite{gierasch85} approximation to maintain a level of similarity with existing 1D studies \citep[e.g.][]{ackerman01,gao18a}, and will investigate the complexities of including both the effects of large scale circulation and sub--grid processes on cloud vertical transport, in a future study. 

\begin {table}
\begin{center}
\begin{tabular}{ l|c|c }
{\bf{Condensate Species}} & {\bf{$q_{\textrm{t}}^{\textrm{below}}$ (g/g)}}\\
\hline
NH$_3$ & 4.48 $\times$ 10$^{-4}$ \\
Fe & 4.48 $\times$ 10$^{-4}$ \\
KCl & 6.10 $\times$ 10$^{-6}$ \\
MgSiO$_3$ & 1.55 $\times$ 10$^{-3}$ \\
Mg$_{2}$SiO$_{4}$ & 1.09 $\times$ 10$^{-3}$ \\
Al$_{2}$O$_{3}$ & 1.11 $\times$ 10$^{-4}$ \\
Na$_{2}$S & 5.32 $\times$ 10$^{-5}$ \\
MnS & 2.53 $\times$ 10$^{-5}$ \\
Zns & 3.72 $\times$ 10$^{-6}$ \\
Cr & 1.77 $\times$ 10$^{-5}$ \\
\end{tabular}
\vspace{+10pt}
\caption{The sub--cloud mass mixing ratio (g/g) for each condensate species. This value defines the total available condensible gas at the point immediately below the cloud base, and in turn sets an upper limit on the total cloud mass. Values are calculated using ATMO, a 1D radiative-convective equilibrium code \protect{\citep[see][]{tremblin15}} using equilibrium chemistry and assuming solar metallicity.}
\vspace{-10pt}
\label{tab:gasmix}
\end{center}
\end{table}

For each condensate species included in the model, EddySed first determines the location (pressure depth) of the cloud base, defined as $z$ = 0, using condensation curve data. The model sets a total mixing ratio immediately below the cloud base, $q_{\textrm{t}}^{\textrm{below}}$ (see Table \ref{tab:gasmix} for this value for each species) which defines the total available condensible vapour. At the base, after assuming that all supersaturated vapour is condensed out, the total mixing ratio then diminishes as it is turbulently mixed upwards, since the solution for the mixing ratio of the cloud, $q_{\textrm{c}}$ is a balance between vertical up--mixing of both the vapour and condensed cloud, $q_{\textrm{t}}$, and the precipitation of condensed material, $q_{\textrm{c}}$, via the free sedimentation efficiency parameter $\fsed$. No cloud microphysics is considered, and condensates are formed homogeneously via chemical phase--equilibrium with no supersaturation assumed beyond the condensation point. This parameterised method allows for fast computation, highly advantageous within a 3D GCM. Particle sizes follow an assumed lognormal size distribution, the geometric standard deviation of which is given in Table \ref{tab:params}. The effective particle radius is defined through the integration of this applied distribution, and the system of equations is analytically closed using a fit of the particle radius dependence of the sedimentation velocity (see Equation 11 in \cite{ackerman01}).

In \citet{lines18a}, we used the model described in \citet{helling13} and previously used by \citet{lee16} to calculate cloud properties on a cell--by--cell basis. However, due to its parameterised vertical structure via the balancing of mixing processes and precipitation, EddySed solves for a single atmospheric column using an input temperature--pressure profile, global effective temperature and surface gravity. After each hydrodynamical time step, we therefore called an interface module which loops through each latitude ($\phi$ = 90 cells) and longitude ($\lambda$ = 144 cells), obtaining the pressures and temperatures for each vertical level and then executed the main EddySed routines which calculate cloud properties for each cell in that column. For computational efficiency, we chose to obtain the scattering and absorption coefficients via a lookup table of the Mie coefficients, rather than performing a numerical integration of these values using a direct implementation of Mie theory at runtime. The cloud radiative coefficients were calculated for each of the condensate, radius and wavelength bins, and amalgamated for a combined opacity.

Our EddySed--UM coupled model does not consider horizontal advection of cloud (vertical advection and turbulent mixing are parameterised via Eqn. \ref{kzz}). Therefore, the cloud chemical and physical properties (condensate mixing ratios and effective radii) were returned to the UM for both `diagnostic' output (i.e. where the quantity does not impact the evolution of the simulation), and for calculation of cloud radiative properties (opacity, single scattering albedo and asymmetry parameter) only\footnote{No significant changes are made to the standalone EddySed code; we interface between EddySed and the UM via an `exchange' module which passes PT profiles to EddySed and cloud properties from it.}. Heating rates, due to scattering and absorption from both gas and cloud, are calculated by SOCRATES and fed back into the dynamical evolution of the model. The number of EddySed wavelength bins (192) exceeds that which we employ in SOCRATES \citep[where 32 bands has been shown to be sufficient for our previous 3D GCM simulations, see][]{amundsen14}, therefore the EddySed opacities are averaged onto the SOCRATES band structure. Aside from the heating rates, SOCRATES, can also be used to calculate synthetic observations \citep[transmission and emission spectra and phase curves, see for example,][]{amundsen16,drummond18b,drummond18c,lines18a,lines18b}. This calculation is done by a further diagnostic call to the radiation scheme at much higher resolution (for this work 500 bands), which exceeds that used by EddySed, requiring interpolation of the cloud properties onto this band structure.

\subsection{Simulations}
\label{subsec:sims}

We performed a total of four simulations all of which are initialised similarly to those performed by \citet{lines18a}, using the `spun--up' or equilibrated, results of previous cloud--free simulations of HD~209458b. We adopted two different simulations as our starting points, one initialised using a temperature--pressure profile from the 1D radiative--convective code ATMO \citep[see][]{tremblin15}, termed the standard deep interior or SDI, and another which was initialised using a global increase of 800\,K of the temperature, termed the hot deep interior or HDI designed to mimic the steady--state solution of \citet{tremblin17} \citep[see][for details]{amundsen16}. The SDI and HDI cloud--free simulations have both reached a quasi--steady state, indicated by cessation of significant evolution in wind velocities in the upper atmosphere, and have run for 1200 and 800\,days (hereafter days refers to Earth days), respectively. 

For each SDI and HDI setup we then performed two simulations with differing values of the cloud sedimentation efficiency factor, $\fsed$ = 0.1 and $\fsed$ = 1.0, which correspond to (and is sometimes refered to) as extended and compact cloud respectively. The value of $\fsed$ can somewhat be informed from fitting to observations with values derived of up to 10 for brown dwarf studies \citep{saumon08}, where cloud is expected to settle below the photosphere across the L--T transition, to as low as 0.01 for super--Earths \citep{morley15} where highly lofted cloud is required to match the observed flat spectra. Since the atmospheres of hot Jupiters are dynamically active and their circulation is expected to drive strong vertical mixing \citep{zhang18,menou18}, in this study we choose a value of $\fsed$ = 0.1 to reproduce the potentially suspended clouds from upwards vertical transport. However, given the uncertainties in the ability for large condensate particles to remain lofted \citep{parmentier13}, particularly on the nightside, we also consider the more compact clouds generated given $\fsed$ = 1.0 that could represent more quiescent atmospheric conditions, as well as addressing the poor constraints on cloud particle sizes.

All four simulations, i.e. SDI with $\fsed$ = 0.1 and $\fsed$ = 1.0 and HDI with $\fsed$ = 0.1 and $\fsed$ = 1.0, were run for a total of 500\,days, with the cloud opacity initially reduced by the cloud opacity scaling factor in Table \ref{tab:params}, for stability reasons, for the first 100\,days only. The addition of a significant cloud opacity to our previously simulated cloud--free atmosphere results in a departure from the previous equilibrium to a new cloudy equilibrium state. Therefore, the heating rates that arise due to the cloud presence can cause the UM to become numerically unstable, as experienced in \citet{lines18a} and \citet{roman18}. To avoid this issue, and similar to \citet{lee16}, we initially reduced the cloud opacity by a factor of 100 and ramped it back to full opacity, linearly, over a 100\,day period, a small fraction of our total 500 simulated days. Additionally, to assist in reducing possible transients and oscillations that arise from the strong cloud opacity feedback, we implemented a three--point boxcar averaging, both horizontally and vertically, to the cloud opacity. Throughout the simulations, we used the vertical damping and diffusion schemes from \citet{lines18a} with the coefficients and damping geometry inline with our previous work and given in Table \ref{tab:params}. The damping parameters remained fixed over the duration of the simulations, unlike \citet{lines18a} where atmospheric instabilities necessitated their increase.

We included ten condensate species in each of our simulations, the full list shown in Table \ref{tab:params} and their corresponding sub--cloud mass mixing ratios are given in Table \ref{tab:gasmix}. The values of $q_{\textrm{t}}^{\textrm{below}}$ are obtained from the 1D radiative--convective equilibrium model ATMO \citep[see][]{tremblin15} assuming a solar metallicity HD~209458b which includes equilibrium gas--phase chemistry. For the SDI simulations, the cloud base solution for high--temperature species is found at higher pressures than our simulation bottom boundary of 2.0 $\times$ 10$^5$ mbar. In this case, we forced such condensates to form their base in our lower most vertical level by feeding a high `dummy' temperature to the EddySed routines at the lower boundary. We therefore caution and acknowledge that the results from our SDI simulations may well overestimate the cloud abundances throughout the atmosphere and the condensate mixing ratios for high--temperatures species should be seen as an upper limit.

As we were interested in the effects of the cloud radiative feedback we derived synthetic observations at both the beginning and end of our simulations i.e., t = 0 and 500\,days, including full orbit phase curves derived as the simulation evolves through a complete orbital period. These observations allow us to compare the impact of fully equilibrated, radiatively active cloud structures against a cloud--free simulated atmosphere.


\section{Results $\&$ Discussion}
\label{sec:resultsdis}

In this section we introduce and discuss the key results from our simulations. Firstly, we show that the impact of cloud radiative feedback is vital to include in such simulations, in Section \ref{subsec:rad_clouds}. Secondly, in Section \ref{subsec:cloud_prop} we discuss how the cloud properties strongly depend on the adopted deep atmosphere temperature--pressure profile, and sedimentation parameter, as well as being significantly different to our previous results obtained using a microphysical cloud model \citep{lines18a,lines18b}. Finally, we show an improved match with the observations of our simulations obtained using the EddySed code, over our previous cloud--free or cloudy simulations in Section \ref{subsec:synth_obs}. 

\subsection{Radiative Impact Of Clouds}
\label{subsec:rad_clouds}

\subsubsection{All Simulations: General Trend}
\label{subsubsec:gen_trend}

\begin{figure*}
\includegraphics[scale = 0.7, angle = 0]{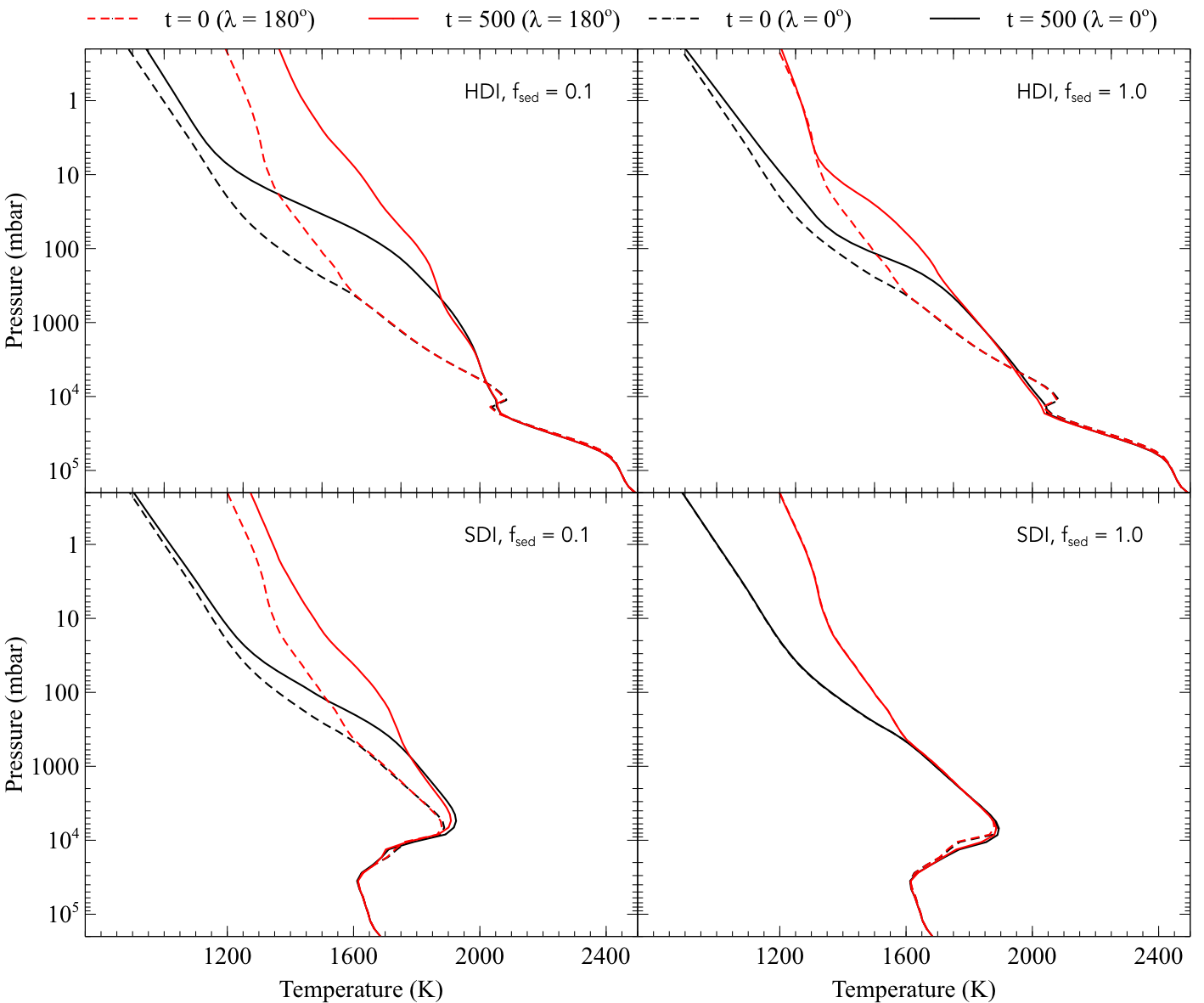}
\caption{Equatorial temperature--pressure profiles, sampled at t = 0 (dashed lines) and t = 500 (solid lines)\,days and for the dayside sub--stellar point, $\lambda$ = 180$^{\circ}$ (red lines) and nightside anti--stellar point, $\lambda$ = 0$^{\circ}$ (black lines) for all simulations.}
\label{fig:all_temp_1d}
\vspace{-10pt}
\end{figure*}

To explore the critical impact of including radiative feedback from cloud opacity, we can simply compare the atmospheric thermal structure and cloud properties at t = 0 against the simulation at t = 500\,days. Since the initial state at t = 0 is an equilibrated cloud--free simulation, the cloud structures that form in the first call to the EddySed routine match those one would expect when either neglecting cloud radiative feedback, or post--processing cloud structures from clear skies simulations \citep[as performed by, for example,][]{parmentier16}. Figure \ref{fig:all_temp_1d} shows the equatorial temperature as a function of pressure for all our simulations at the start (t = 0\,days, dashed lines) and end (t = 500\,days, solid lines), at the sub--stellar point, $\lambda$ = 180$^{\circ}$ (red lines) and anti--stellar point, $\lambda$ = 0$^{\circ}$ (black lines).

\begin{figure*}
\includegraphics[scale = 0.7, angle = 0]{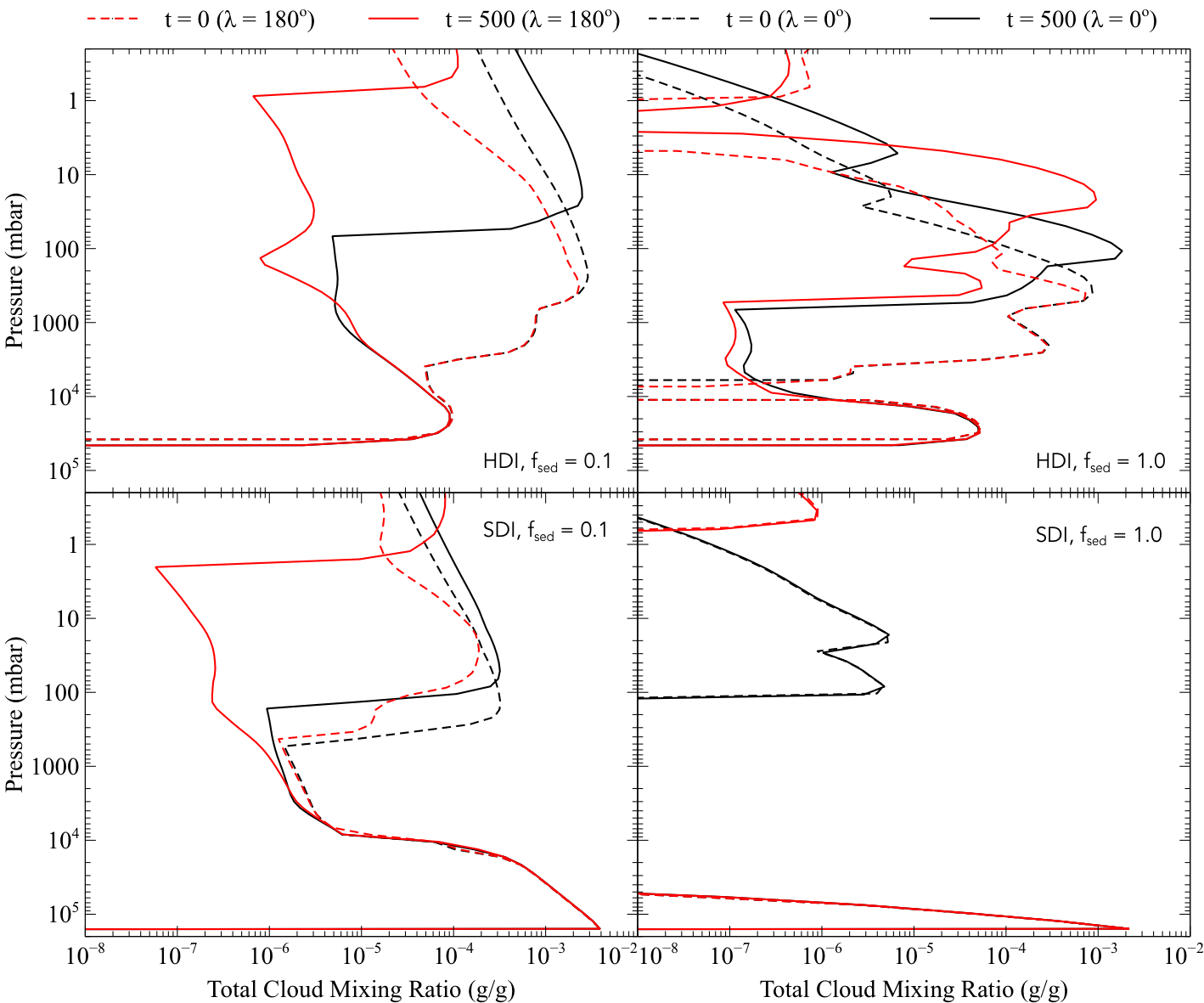}
\caption{Equatorial total cloud mixing ratio profiles, sampled at t = 0 (dashed lines) and t = 500 (solid lines)\,days and for the dayside, $\lambda$ = 180$^{\circ}$ (red lines) and nightside, $\lambda$ = 0$^{\circ}$ (black lines) for all simulations.}
\label{fig:all_cloud_1d}
\end{figure*}

The temperature--pressure profile evolves quickly over the first $\sim$ 100\,days (not shown), and is steady thereafter; evolution of the atmosphere towards its equilibrium state with radiatively--active clouds would occur even faster, but is controlled by the stability--enhancing opacity ramping over this initial 100 day period. For all but the standard deep interior and compact cloud simulation (SDI $\fsed$ = 1.0, bottom right panel), there is a substantial change in the atmosphere's thermal structure. At the equator, both night and day hemispheres see an increase in temperature for almost all pressure depths; the high velocity jet at the equator efficiently re--distributes heat onto the nightside causing the temperatures at the anti--stellar point to rise by up to 300\,K. However, at the sub--stellar point, temperature increases are typically larger than those on the nightside, resulting in an increased day--night temperature contrast for simulations including cloud radiative feedback.

The temperature changes are caused by the heating (or cooling) of the atmosphere induced by the cloud opacity. Clouds can influence the atmospheric temperature via a number of mechanisms. Outgoing planetary thermal emission can be trapped by an optically thick cloud base causing localised heating beneath a cloud structure. In \cite{lines18a}, this effect can be seen in the full orbit thermal phase curves via the advection--driven modulation of the outgoing thermal flux. Cloud particles can also absorb stellar photons. Depending on the location of the cloud top, this heating due to cloud absorption can restrict the stellar insolation from penetrating deeper into the atmosphere. Finally clouds can reflect (backscatter) photons which can reduce absorption (from both cloud particles and gas species) leading to a cooling atmosphere. To explore the effect of these temperature changes further, we show, in Figure \ref{fig:all_cloud_1d}, the total cloud mixing ratio as a function of pressure, for all our simulations at 0 and after 500\,days, at the equator but for two longitudinal positions: the anti--stellar, $\lambda$ = 0$^{\circ}$ (black lines) and sub--stellar, $\lambda$ = 180$^{\circ}$ (red lines) points. The total cloud mixing ratio considers a summation of the mixing ratios from all present condensates; we take a specific look at the distribution and role of each individual condensate species in Section \ref{subsubsec:hdi}. The differences between the various simulations are explored in more detail in Section \ref{subsec:cloud_prop}, but for now Figure \ref{fig:all_cloud_1d} shows the dramatic change in total cloud abundance when cloud radiative feedback in included. While the SDI and $\fsed$ = 1.0  simulation (bottom right panel) shows little change in the distribution of cloud, due to the strong link between the temperature--pressure and the formation of phase--equilibrium cloud, all other simulations show a striking change in the cloud structure, when including cloud radiative feedback.

Figures \ref{fig:all_temp_1d} and \ref{fig:all_cloud_1d} together reveal that where a large change in the temperature--pressure structure exists, there is a corresponding difference in the cloud mixing ratio, as one might expect. Typically, radiative heating from clouds causes the cloud mixing ratio to decrease at deeper pressures and increase at lower pressures; effectively a significant fraction of the cloud in a given column elevates to lower pressures due to the condensation inhibiting temperatures extending further, vertically, into the upper atmosphere. Overall, the cloud mixing ratio is influenced by two thermally--driven mechanisms: the formation of a cloud condensate base at lower, cooler pressures (see Section \ref{subsubsec:hdi}), and the relation of the cloud mass with the eddy diffusivity and mixing length, both of which are dependent on temperature. While our simulations with cloud radiative feedback show a change in cloud abundance for both the day and night sides, the change in the cloud mixing ratio is more pronounced at the sub--stellar point, explained by the aforementioned trend in larger temperature changes on the dayside hemisphere which occur due to the lack of direct stellar heating on the nightside.

\begin{figure*}
\includegraphics[scale = 0.65, angle = 0]{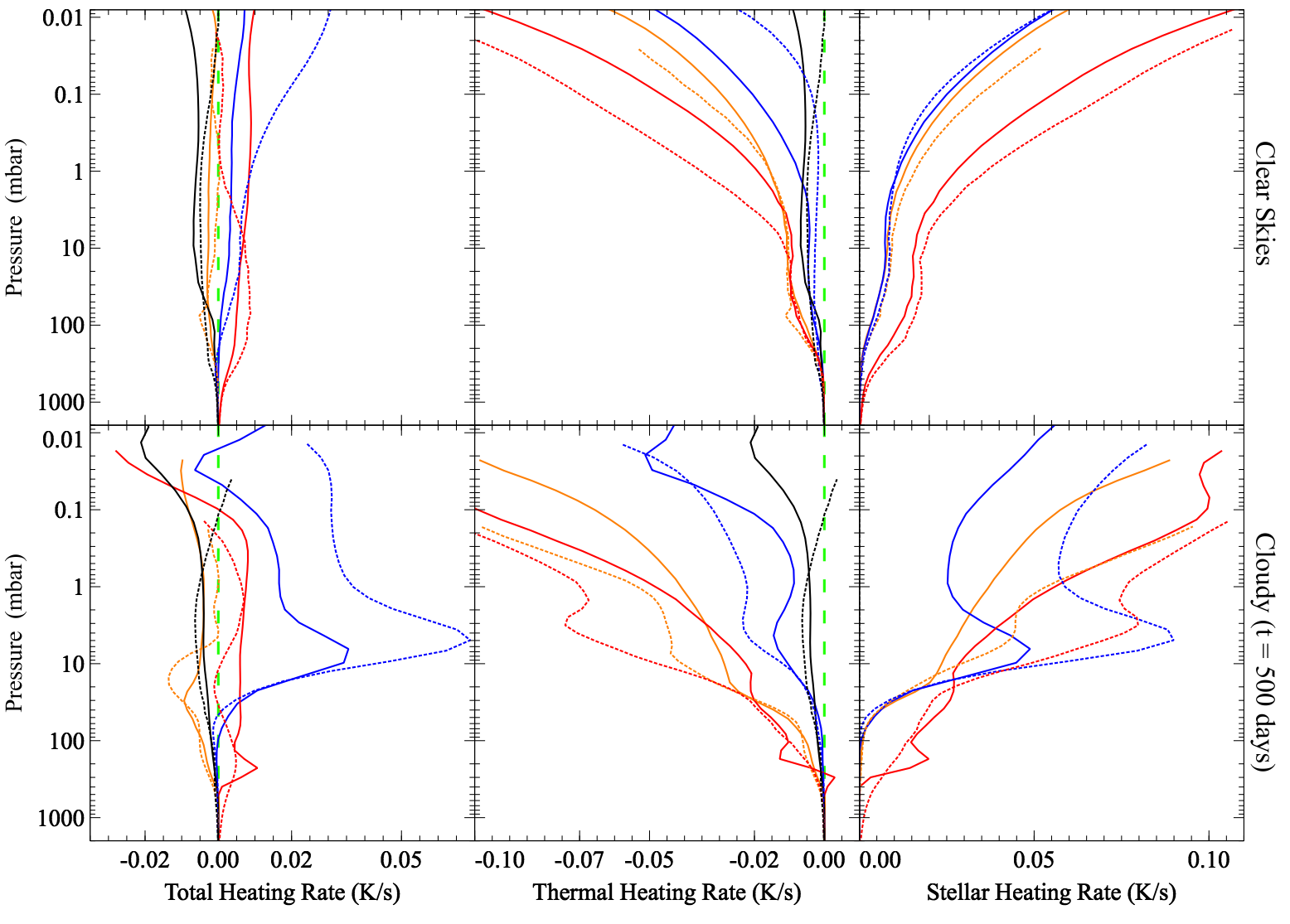}
\caption{Total, thermal (long--wave) and stellar (short--wave) heating rates for a clear sky (upper) and evolved cloudy (hot deep interior and $\fsed$ = 0.1 case) atmosphere at t = 500\,days (lower) for the equator $\phi$ = 0$^{\circ}$ (dotted lines) and mid--latitudes $\phi$ = 45$^{\circ}$ (solid), at the anti--stellar, $\lambda$ = 0$^{\circ}$ (black lines) and sub--stellar, $\lambda$ = 180$^{\circ}$ (red lines) points, as well as the east--limb, $\lambda$ = 260$^{\circ}$ (orange lines) and west--limb, $\lambda$ = 100$^{\circ}$ (blue lines). The dashed green lines indicate the zero heating rate.}
\label{fig:hdi_f01_hr_1d}
\end{figure*}

\begin{figure*}
\begin{subfigure}{0.48\textwidth}
\includegraphics[scale = 0.12, angle = 0]{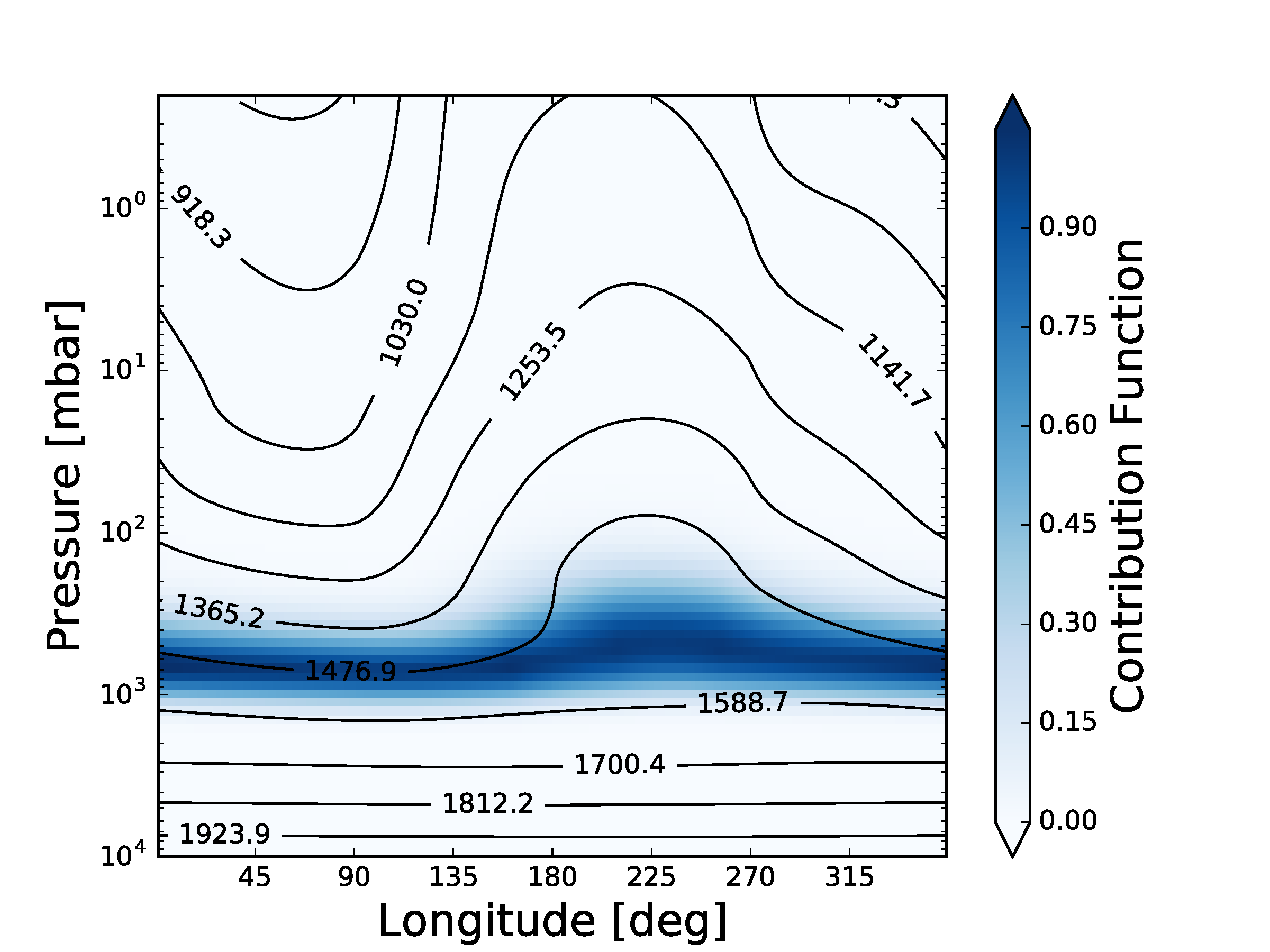}
\end{subfigure}\hspace*{\fill}
\begin{subfigure}{0.48\textwidth}
\includegraphics[scale = 0.12, angle = 0]{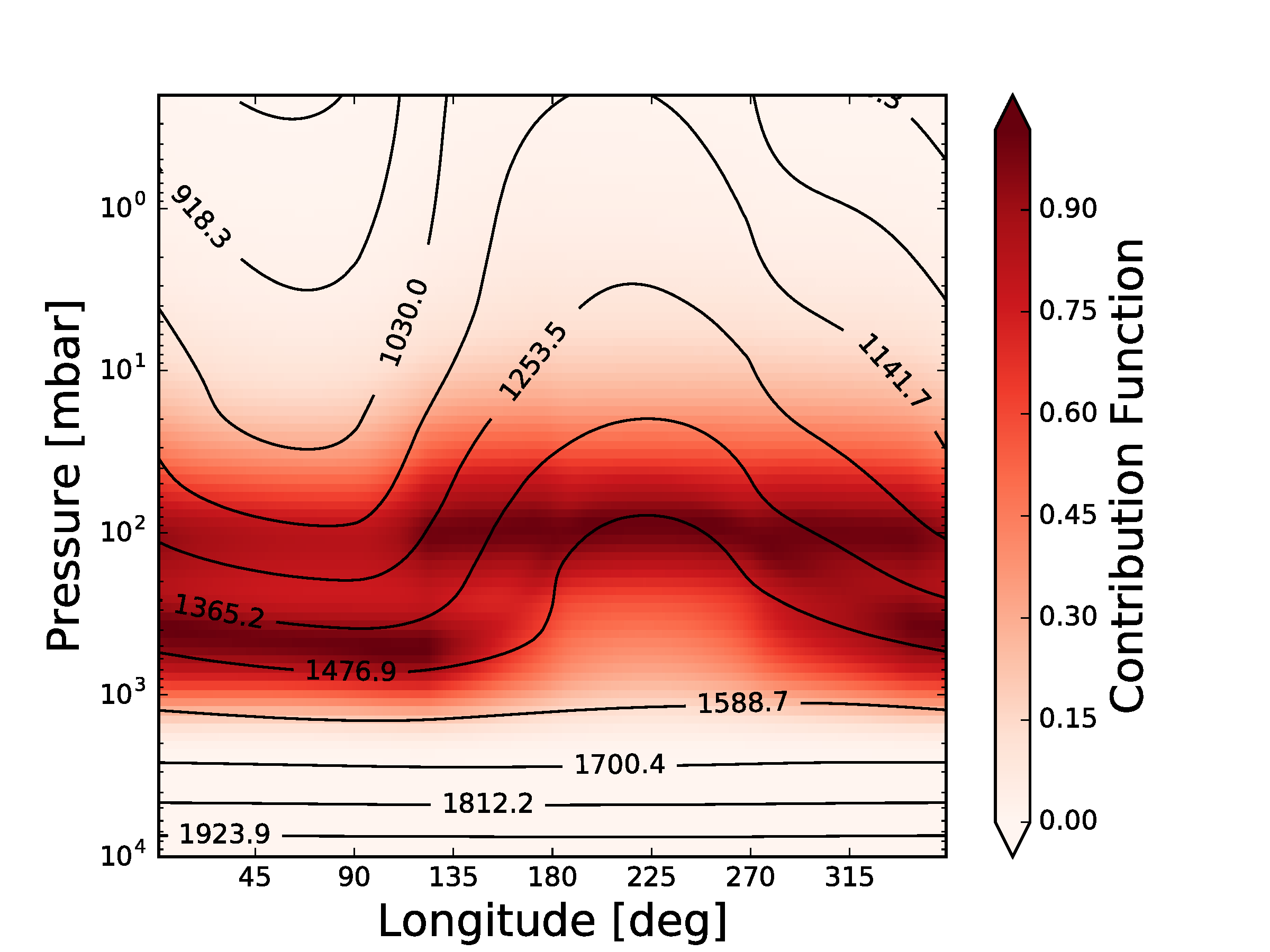}
\end{subfigure}
\medskip
\begin{subfigure}{0.48\textwidth}
\includegraphics[scale = 0.12, angle = 0]{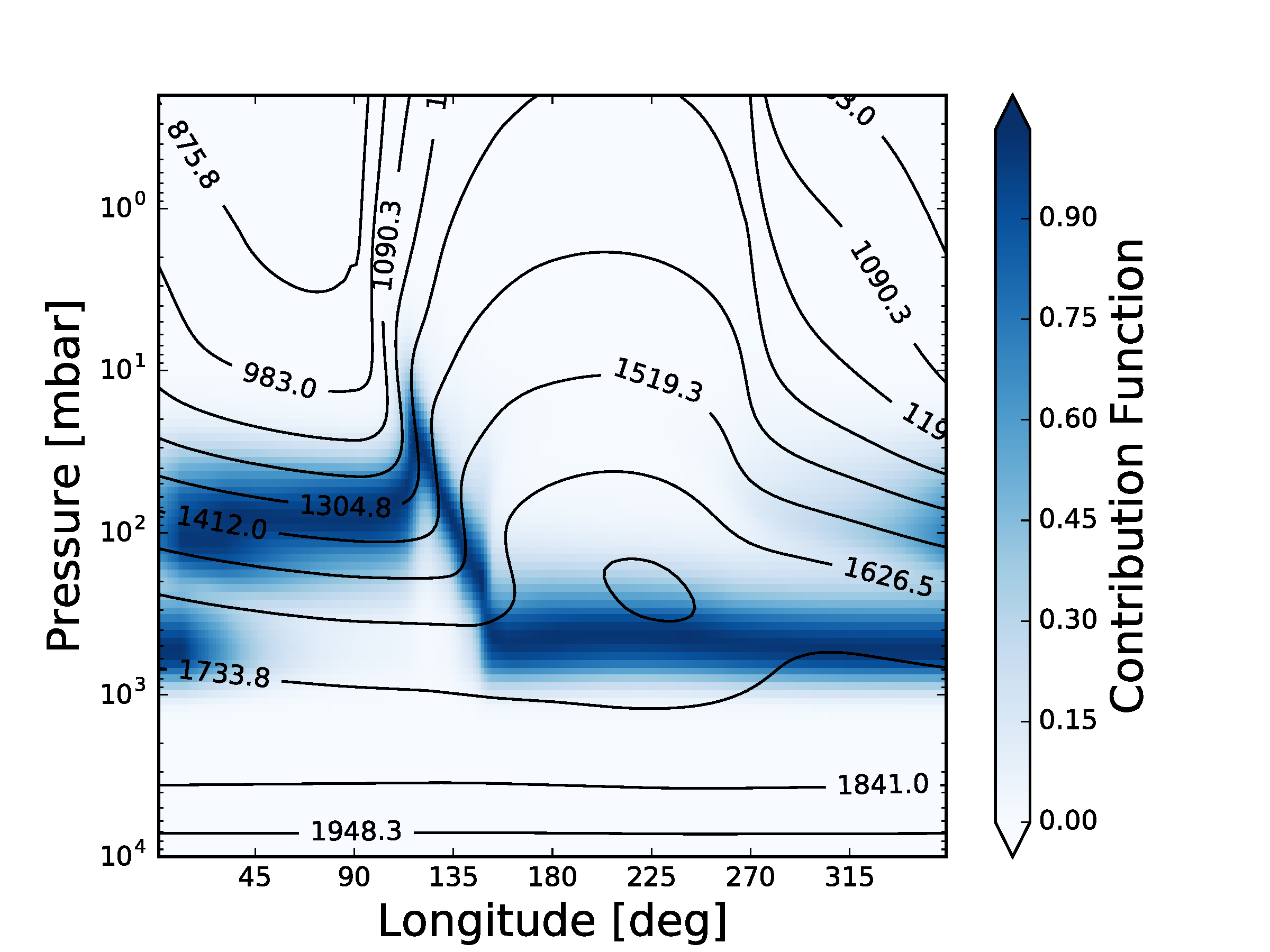}
\end{subfigure}\hspace*{\fill}
\begin{subfigure}{0.48\textwidth}
\includegraphics[scale = 0.12, angle = 0]{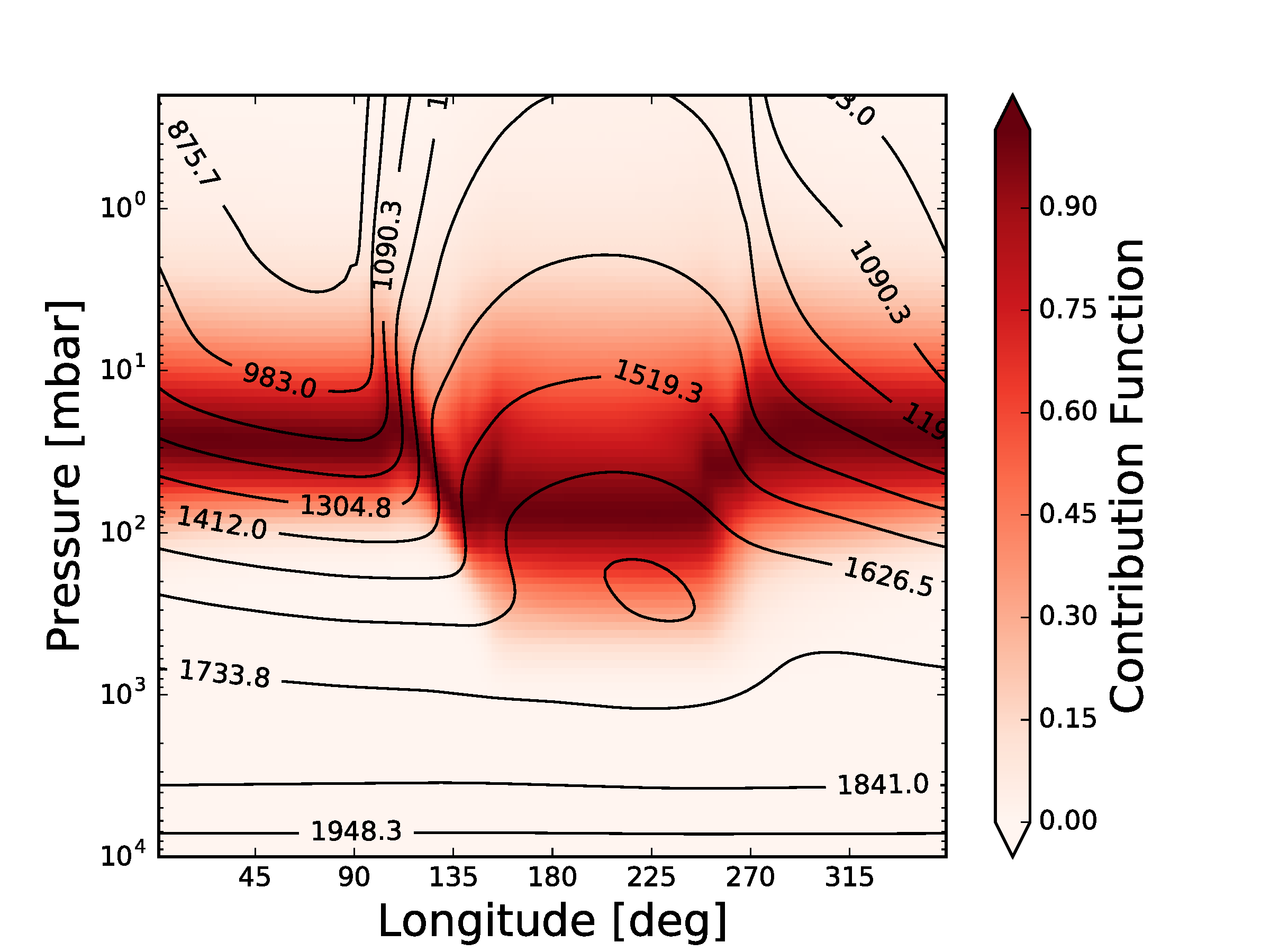}
\end{subfigure}
\caption{Meridional-mean of the normalised contribution function (colour scale) for clear sky (upper row) and cloudy HDI and $\fsed$ = 0.1 (lower row) for 0.5\,$\mu$m (left column) and 4.5\,$\mu$m (right column). Atmospheric temperature is shown via black contours.}
\label{fig:cf}
\end{figure*}

To better understand the effect of radiatively active cloud on the atmospheric temperature, in Figure \ref{fig:hdi_f01_hr_1d} we present, for a clear sky (upper) and the hot deep interior and extended (HDI $\fsed$ = 0.1) simulation at t = 500\,days (lower), the total, thermal (long--wave) and stellar (short--wave) heating rates at the equator $\phi$ = 0$^{\circ}$ (dotted lines) and at mid--latitudes $\phi$ = 45$^{\circ}$ (solid lines), and for four longitudes: the anti--stellar and sub--stellar points and east--limb and west--limb, $\lambda$ = 0$^{\circ}$ (black lines), 180$^{\circ}$ (red lines), 260$^{\circ}$ (orange lines) and 100$^{\circ}$ (blue lines) respectively. The dashed green lines indicate the zero heating rate. While the full 3D implications of the cloud properties, temperatures and heating rates are discussed in Section \ref{subsubsec:hdi}, we highlight for now that the presence of cloud on the dayside can, compared to a clear-sky atmosphere, result in large positive net heating rates for the majority of the upper atmosphere ($<$ 1000\,mbar), whereas the nightside and east--limb contributes a cooling due to cloud radiative--emission.

Figure \ref{fig:cf} shows the meridional average of the normalised contribution function for a clear sky (upper) and cloudy (t = 500\,days) HDI and $\fsed$ = 0.1 atmosphere (lower) for both 0.5\,$\mu$m (left) and 4.5\,$\mu$m (right). We obtain this value using the methodology described in \cite{drummond18c}. The peak of the normalised contribution function effectively describes the pressure of the wavelength--dependent photosphere. The location of the thermal (4.5\,$\mu$m) photosphere is shown to rise to lower pressures, due to the presence of cloud; an expected effect due to the opaque nature of the condensate particles. The rising thermal photosphere, to lower pressures, would initially indicate (without further consideration of the PT profile) that the overall equilibrium temperature of the planet reduces, in response to a decreasing temperature with altitude, despite the atmospheric temperature increase due to cloud absorption. However, we find that by analysing the outgoing thermal flux, there is actually an increase in the equilibrium temperature. This occurs because the photosphere is raised to a lower pressure region which is, at the end of the simulation, warmer than at the original (clear sky) pressure level of the photosphere; this is a product of the intense atmospheric heating by clouds. This increase in planetary equilibrium temperature is therefore consistent with both the calculated contribution function and the thermally evolved PT profile due to the cloud absorption. The ability for the photosphere to increase in altitude, but also return an increased thermal flux due to heating, is also found in \cite{drummond16} in their gas--phase chemistry study.

\subsubsection{Hot Deep Interior $\&$ $\fsed$ = 0.1 Simulation}
\label{subsubsec:hdi}

In order to provide a more detailed look into the importance of cloud radiative feedback, in addition to a comprehensive analysis of the cloud properties, we focus on one simulation as representative of the main effects found in our simulation set. We choose the hot deep interior, HDI, with $\fsed$ = 0.1 for two reasons. Firstly, previous modelling of hot Jupiter atmospheres with EddySed indicates that $\fsed$ = 0.1 is more reflective of their atmospheric conditions and, therefore, it is a more physically motivated choice. Secondly, we explore in detail a hotter atmosphere in \citet{lines18a}, matching the HDI initial conditions, and therefore this simulation provides a fairer comparison when considering the differences arising between parameterised and microphysics models, although as mentioned earlier a full comparison requires a full matching of model input parameters, such as the condensate lists included in both models.

\begin{figure*}
\includegraphics[scale = 0.65, angle = 0]{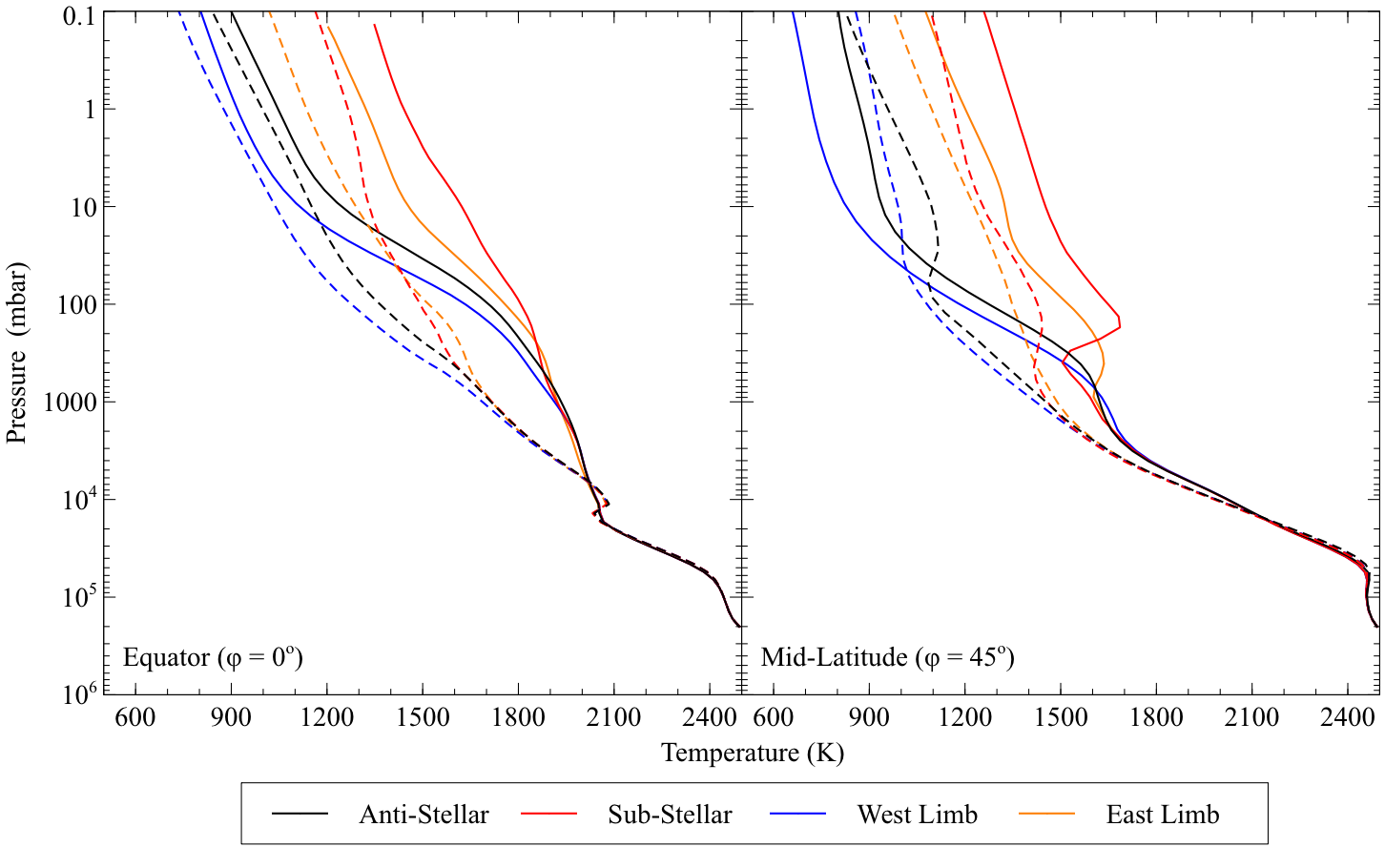}
\caption{Temperature--Pressure profiles, sampled at t = 0\,days (dashed lines) and t = 500\,days (solid lines) for the equator (left panel) and mid--latitude (right panel) at the anti--stellar, $\lambda$ = 0$^{\circ}$ (black lines) and sub--stellar, $\lambda$ = 180$^{\circ}$ (red lines) points, as well as the east--limb, $\lambda$ = 270$^{\circ}$ (orange lines) and west--limb, $\lambda$ = 90$^{\circ}$ (blue lines), for the hot deep interior and $\fsed$ = 0.1 case.}
\label{fig:hdi_f01_temp_1d}
\end{figure*}

Figure \ref{fig:hdi_f01_temp_1d} shows the temperature--pressure profiles for the HDI $\fsed$ = 0.1 simulation at t = 0 (dashed lines) and t = 500 (solid lines)\,days for both the equator, $\phi$ = 0$^{\circ}$ (left) and mid--latitude, $\phi$ = 45$^{\circ}$ (right) for four key longitudes: the anti--stellar, $\lambda$ = 0$^{\circ}$ (black lines) and sub--stellar, $\lambda$ = 180$^{\circ}$ (red lines) points and the east--limb, $\phi$ = 270$^{\circ}$ (orange lines) and west--limb, $\phi$ = 45$^{\circ}$ (blue lines). Crucially, by exploring additional zonal positions, over those shown in Figure \ref{fig:all_temp_1d}, we can see that not all locations in the atmosphere undergo a temperature increase, and that the change in temperature is also dependent on latitude. While all longitudes report a temperature increase at the equator (for most pressures), at higher latitudes both the anti--stellar and west--limb see a cooling for pressures lower than 30\,mbar. Hence, combined with heating at the sub--stellar point, for the low pressures found in the upper atmosphere, large temperature contrasts can occur; the maximum temperature contrast (between the west--limb and sub--stellar point) can reach up to 650\,K for higher latitudes. At the equator, the intense dayside heating also drives large temperature contrasts ($\sim$ 500\,K), but the equatorial jet transports this heat to the nightside more efficiently than at mid-latitudes, leading to an overall warming of the nightside equator despite the enhanced nightside cooling from the cloud.

At higher pressures, deeper in the atmosphere; while there is a large cloud--driven heating between 10 and 1000\,mbar at the equator, the temperature increase at corresponding pressures for mid--latitudes, is much less. This is a result of the stellar insolation being able to penetrate further down into the atmosphere at the equator before being absorbed, due the lack of cloud on the dayside caused by condensation--inhibiting high temperatures (while this effect occurs from t = 0 days, it is much more important at t = 500 days since radiative heating expands the cloud--free region for a range of lower--temperature condensates). This suggests caution is required when using a single or averaged 1D temperature--pressure profile to represent an atmosphere which exhibits such zonal and meridional variation. Our result reinforces that of \citet{blecic17}, where retrieval was performed on simulated observations derived from a 3D simulation of a hot--Jupiter, and returned a best--fitting 1D temperature--pressure profile which did not match any profile existing at any spatial location within the simulation, nor did it match the geometric mean profile.

The temperature, for both latitudes shown in Figure \ref{fig:hdi_f01_temp_1d}, shows little variation between t = 0 and t = 500\,days in the deep ($>$ 10$^4$\,mbar) atmosphere, owing in part to the long thermal timescales at these pressures, but also due to the absence of cloud below 4 $\times$ 10$^4$\,mbar. The heating rates in Figure \ref{fig:hdi_f01_hr_1d} indicate that the main driving force behind the atmospheric temperature increase is due to the dayside stellar absorption. The comparison with a clear sky atmosphere shows that when clouds are included, there is a large increase in stellar heating rates for the sub--stellar point, but also for the west--limb where cloud is most abundant due to the cooler temperatures at western longitudes. While the temperature increase for pressures deeper than 300\,mbar (at the point the stellar heating rate has almost fully attenuated) can be driven by the vertical advection of heat generated by stellar absorption, there also exists a component attributed to the absorption of the planetary thermal emission; a small but noticeable spike in the thermal heating rate at around 300mbar indicates this process is operating.

\begin{figure*}
\includegraphics[scale = 0.7, angle = 0]{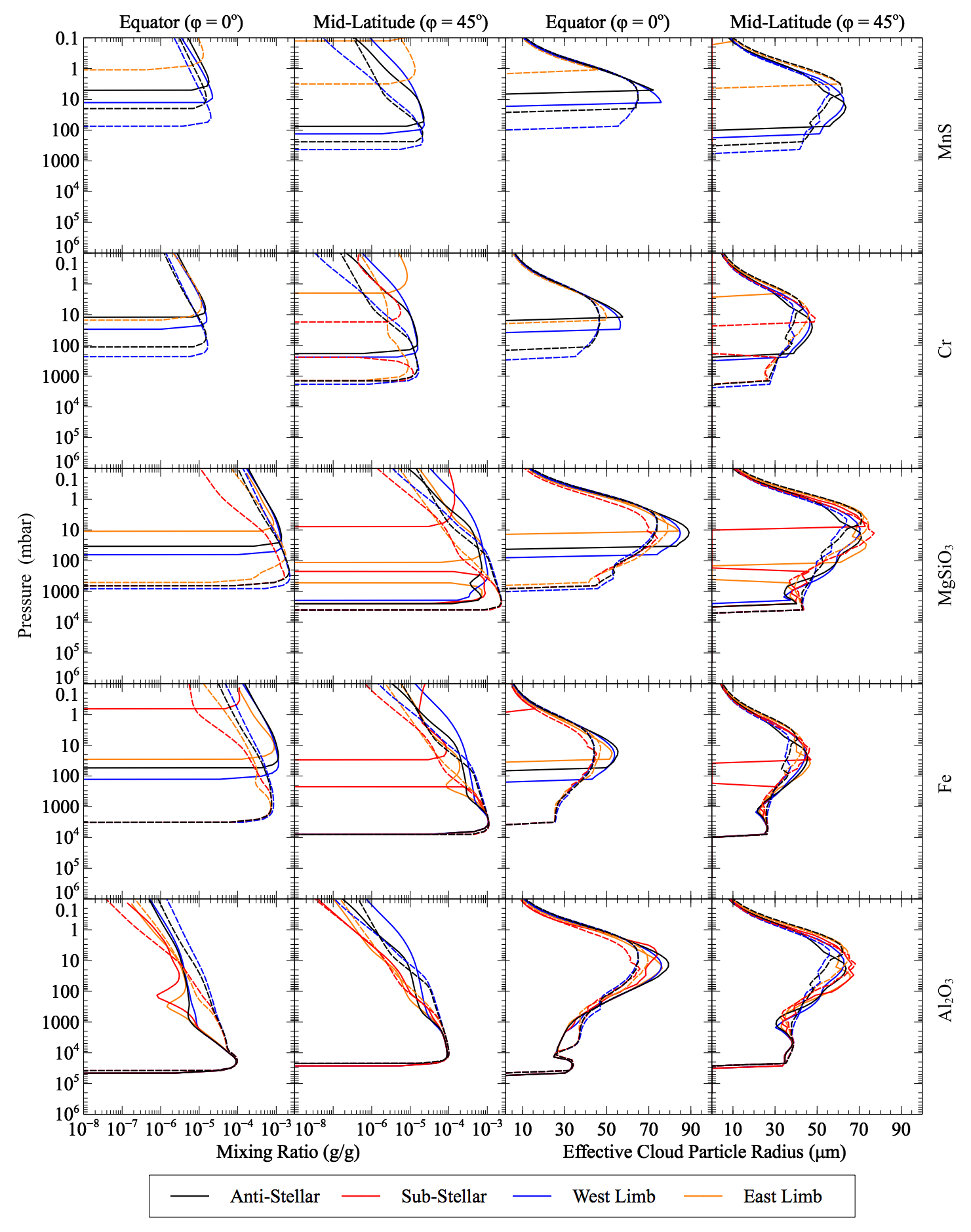}
\caption{Individual condensate mixing ratio and effective radii profiles, sampled at t = 0\,days (dashed lines) and t = 500\,days (solid lines) at the anti--stellar, $\lambda$ = 0$^{\circ}$ (black lines) and sub--stellar, $\lambda$ = 180$^{\circ}$ (red lines) points, as well as the east--limb, $\lambda$ = 270$^{\circ}$ (orange lines) and west--limb, $\lambda$ = 90$^{\circ}$ (blue lines), for the hot deep interior $\fsed$ = 0.1 simulation. The {\textit{total}} cloud mixing ratio typically remains non--zero up to the simulation upper boundary, which varies between 0.1 mbar and 10$^{-3}$ mbar  depending on longitude.}
\label{fig:hdi_f01_mix_1d}
\end{figure*}

\begin{figure}
\includegraphics[scale = 0.5, angle = 0]{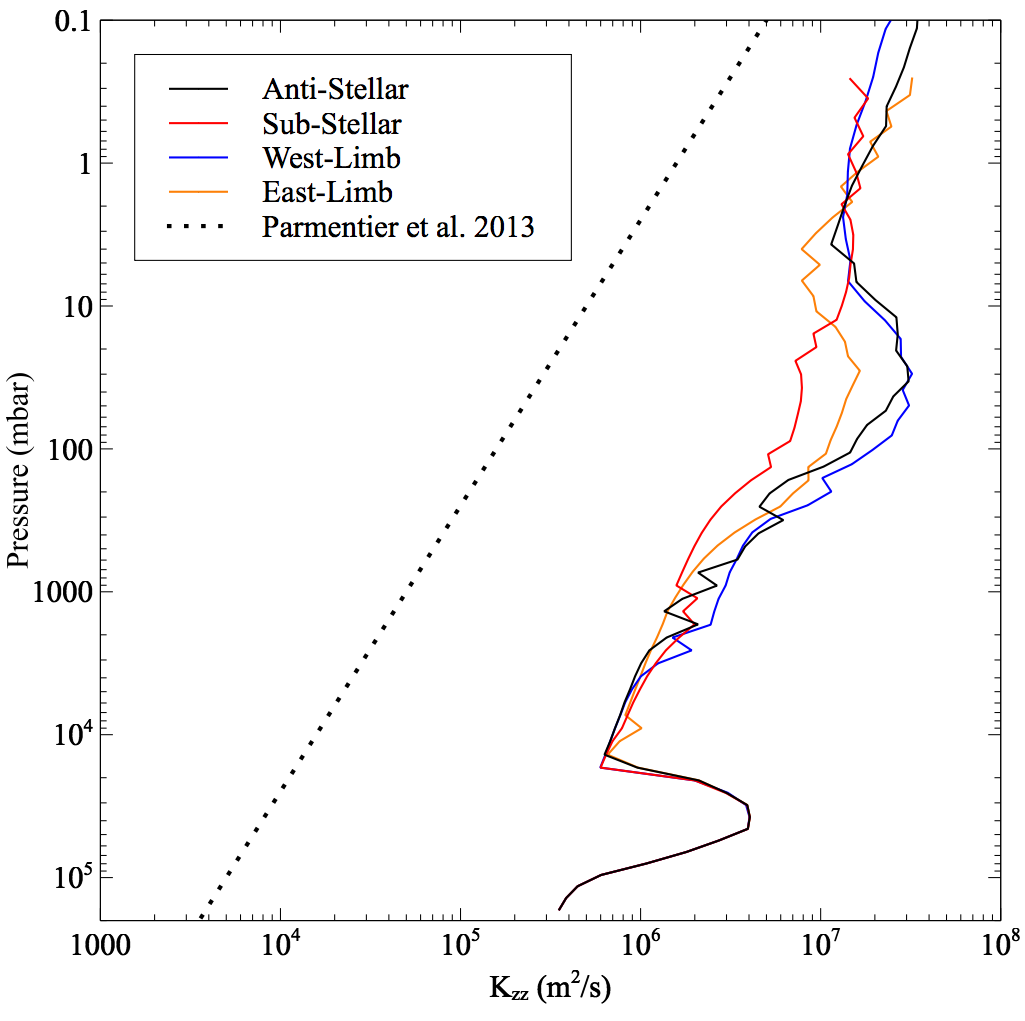}
\caption{$K_{zz}$ profiles, sampled at t = 500\,days at the anti--stellar, $\lambda$ = 0$^{\circ}$ (black lines) and sub--stellar, $\lambda$ = 180$^{\circ}$ (red lines) points, as well as the east--limb, $\lambda$ = 270$^{\circ}$ (orange lines) and west--limb, $\lambda$ = 90$^{\circ}$ (blue lines), for the hot deep interior $\fsed$ = 0.1 simulation. The dotted line follows the analytical form of $K_{zz}$ derived from 3D simulations of tracer advection \protect{\citep{parmentier13}.}}
\label{fig:hdi_f01_kzz_1d}
\end{figure}

The temperature change caused by the radiative impact of clouds can be better understood when analysing it against the geometric distribution of cloud. In Figure \ref{fig:hdi_f01_mix_1d} we show the mixing ratios (first two columns) and effective radii (third and fourth columns) as a function of pressure for our five most abundant condensates (MnS, Cr, $\mso$, Fe and $\alu$) from our HDI $\fsed$ = 0.1 simulation, at the equator, $\phi$ = 0$^{\circ}$ (left) and mid--latitude, $\phi$ = 45$^{\circ}$ (right) and at four key longitudes: the anti--stellar, $\lambda$ = 0$^{\circ}$ (black lines) and sub--stellar, $\lambda$ = 180$^{\circ}$ (red lines) points, and the east--limb, $\phi$ = 270$^{\circ}$ (orange lines) and west--limb, $\phi$ = 45$^{\circ}$ (blue lines). We do not include Na$_2$S in our figures due to its negligible contribution (Na$_2$S is only present in in a very spatially limited area of our atmosphere), and we choose to not plot $\msotwo$ due to its similar (identical for a range of pressures) condensation curves to $\mso$. NH$_3$, KCl and ZnS do not form since their condensation thresholds are not crossed anywhere in the simulated atmosphere.

It is no surprise that the condensate cloud bases for different species form at different pressures, as shown in Figure \ref{fig:hdi_f01_mix_1d}, since the solution for the cloud lower boundary is obtained from the condensate curve data which is highly dependent on the physical properties of a given condensate. Simply, the higher temperature condensates such as $\alu$ and Fe (bottom two rows of Figure \ref{fig:hdi_f01_mix_1d}) are able to withstand the extreme temperatures of the deep atmosphere, and form at much higher pressures. The change in the cloud base pressure after cloud radiative feedback is included, is sensitive to the change of the local temperatures and pressures caused by the cloud radiative feedback itself. With the exception of $\alu$, with its base occurring in a region of the atmosphere with little temperature change, condensates see an elevation in their cloud bases due to a warming atmosphere from the presence of radiatively active cloud. The shift in cloud base pressure is also dependent upon both longitude and latitude, but can typically span up to one to two orders of magnitude. Where the most substantial heating occurs, such as at the equator and eastern--limb, the temperature increase caused by the radiative feedback of the cloud can lead to the complete evaporation of (or rather, the inhibited condensation of) MnS, Cr and even the more refractory silicates (see top two rows of Figure \ref{fig:hdi_f01_mix_1d}). We have already shown via the 1D temperature--pressure profiles in Figure \ref{fig:hdi_f01_temp_1d} that there can be a drastic change in atmospheric temperature due to the effect of cloud radiative feedback. In the most extreme cases we find that the temperature can shift by up to 300\,K. Such a radical change in temperature is enough to transition across multiple condensation curve thresholds. For example, an absolute change in 300\,K at 100\,mbar {\emph{can}} result in the evaporation of four condensates: Cr, $\mso$, $\msotwo$ and Fe. Indeed, in our precise case, the sub--stellar region does exhibit a complete absence of those four condensates.

The incidence of atmospheric conditions which lead to full evaporation of the cloud, or inhibit its initial formation, is largely dependent on latitude, as the weakened heating effect at higher latitudes can still allow dayside cloud formation in the upper atmosphere for our most volatile species. In certain circumstances, such as for mid--latitude $\mso$ and Fe, localised extreme temperatures (from cloud heating) can inhibit cloud formation over a limited range of pressures. This can allow the formation of multiple cloud decks (for a single species) which, in turn, cause sharp changes in the heating rate as a function of pressure. For example, the opacity window from a Fe free region at $\sim$ 100\,mbar, followed by a second deeper Fe cloud deck at 200\,mbar leads to a steep opacity gradient \citep[particularly since Fe has such a high opacity; see][for a detailed look at the optical properties of potential exoplanet condensates]{kitzmann18} that results in a small but noticeable spike in the heating rates, via the mid--latitude sub--stellar short--wave heating rate, shown in Figure \ref{fig:hdi_f01_hr_1d}, resulting in the temperature spike at the sub--stellar point shown in the temperature--pressure profile in Figure \ref{fig:hdi_f01_temp_1d}. Since there is such a contrast in the cloud abundance with latitude, the heating rates also show variation between the equator and mid--latitudes. At the west--limb, where we experience the largest radiative adjustment, we find that the equator experiences significantly larger heating rates that mid--latitudes. This result can appear unexpected on first consideration, as the cooler mid--latitudes should introduce a larger cloud abundance and hence higher opacity. The answer lies, again, with the effect of the temperature profile on the vertical cloud distribution. Mid--latitudes are indeed cooler than the equator, and therefore the condensate bases are able to form at deeper pressures. Since the cloud mixing ratio decreases with increasing height (and decreasing pressure) as it is mixed upwards, the value of the cloud abundance is set partially by this cloud base. Since cloud condensate bases tend to form at higher altitudes at the equator, due to the higher temperatures, the total mixing ratio can be larger at the equator than mid--latitudes, for the same pressure. This is demonstrated by the mixing ratio of Fe in Figure \ref{fig:hdi_f01_mix_1d} which is roughly an order of magnitude larger at the equator, for the western--limb, for the lowest pressure plotted. The asymmetry in heating rates, with latitude, may have consequences for the formation of atmospheric jets, but we leave an investigation into the effect of cloud on atmospheric dynamics to future work.

Our general trend of a warming atmosphere, on both day and night hemispheres, leading to a reduction in dayside cloud coverage, is qualitatively similar to the results of \cite{parmentier16} who find allowing clouds to radiatively feedback onto their atmosphere causes a reduction in cloud abundance on the irradiated hemisphere. \cite{roman18} also find that including radiatively active clouds in their 3D simulations causes a large change in both the atmospheric thermal state and cloud distribution. However, while they obtain an enhanced abundance of cloud at higher latitudes, and depletion at the equator, we find this only occurs limited places (low pressures, western--limb) where the local atmospheric temperature decreases. We also do not find evidence of the cyclic behaviour of condensation and evaporation for $\alu$ since our temperature profile does not increase to the $\alu$ evaporation point (except for the deep atmosphere where the $\alu$ base is formed (and where our temperature profile does not evolve).

Figure \ref{fig:hdi_f01_kzz_1d} displays our equatorial $K_{zz}$ values, for the anti--stellar (black line) point, sub--stellar (red line) point, west--limb (blue line) and east--limb (orange line). The empirically derived expression from \cite{parmentier13} for $K_{zz}$ is overlaid as a dotted line. We find that $K_{zz}$ varies depending on zonal location, as expected due to the variation in the PT profiles with longitude and pressure. With the exception of the dayside sub--stellar point, where the PT profile is shown in Figure \ref{fig:hdi_f01_temp_1d} to be distinctly more isothermal, there is a peak in the $K_{zz}$ value between 10 and 100 mbar. This localised enhancement corresponds to the size maximum in the effective cloud particle radius in Figure \ref{fig:hdi_f01_mix_1d}, indicating the relevance of the eddy diffusion on cloud properties. 

Overall we find considerably larger values than those obtained in studies of eddy diffusion in hot--Jupiter atmospheres \citep{parmentier13}. The explanation for this is due to the convective atmosphere regime that the \cite{gierasch85} formulation of $K_{zz}$ is appropriate for. Since a large proportion of the atmospheres of hot--Jupiters are expected to be radiative, values of $K_{zz}$ from \cite{gierasch85} are likely inaccurate. The comparison of our results with \cite{parmentier13}, who explore the precise case of tracer transport for HD~209458b, indicates that we may overestimate $K_{zz}$ throughout the atmosphere, although our results are still in-line with commonly used values of $K_{zz}$ which are obtained through the product of the vertical scale height and the root mean square of the vertical velocity \citep[for example][]{lewis10,moses11}.

The most notable consequence of this enhanced $K_{zz}$ is the increased cloud mass and hence opacity in the upper atmosphere due to the efficiency of upwards vertical transport; the magnitude of $K_{zz}$ is able to suspend even the largest deci--micron sized condensate particles from settling out of the photosphere. The impact of a choice of $K_{zz}$ on cloud properties, such as a potentially reduced cloud-RT coupling from more vertically settled cloud, will be explored in future work by implementing newer and physically informed $K_{zz}$ values for hot--Jupiter atmospheres. For now, our investigation into the effect of the sedimentation parameter (see section \ref{subsubsec:sed}) partly explores this effect.

\begin{figure*}
\begin{subfigure}{0.48\textwidth}
\includegraphics[scale = 0.44, angle = 0]{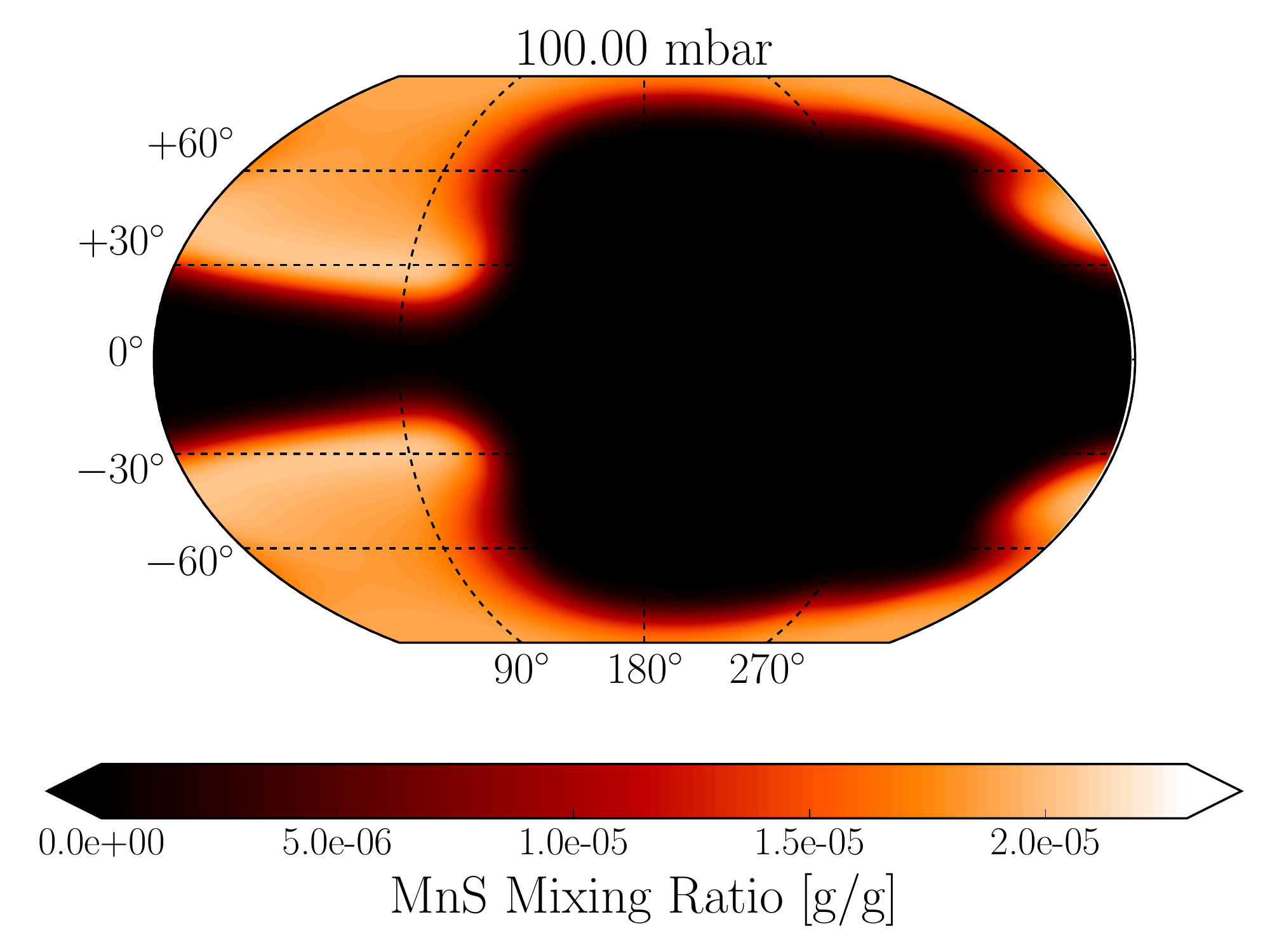}
\caption{}
\end{subfigure}\hspace*{\fill}
\begin{subfigure}{0.48\textwidth}
\includegraphics[scale = 0.44, angle = 0]{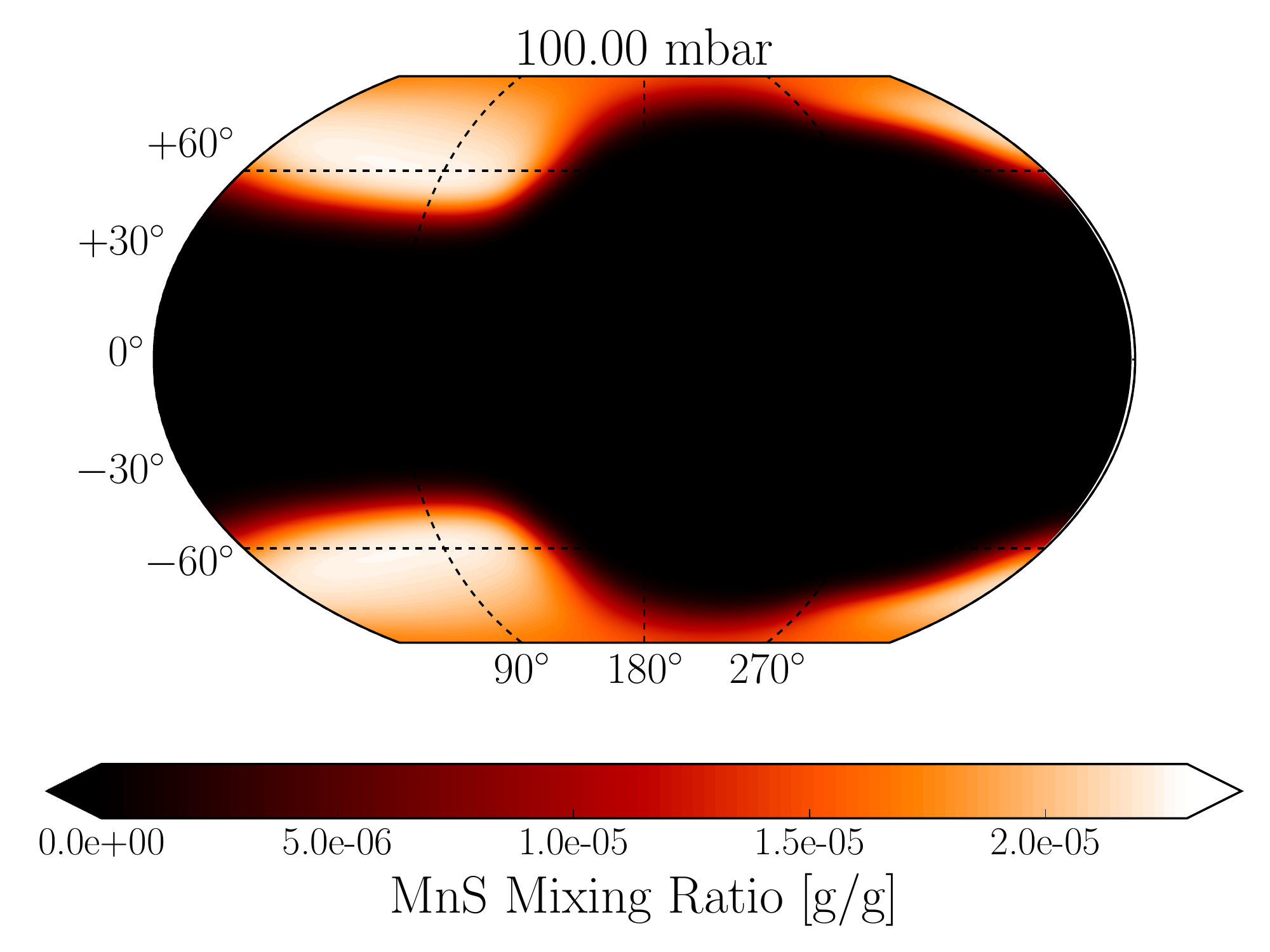}
\caption{}
\end{subfigure}
\medskip
\begin{subfigure}{0.48\textwidth}
\includegraphics[scale = 0.44, angle = 0]{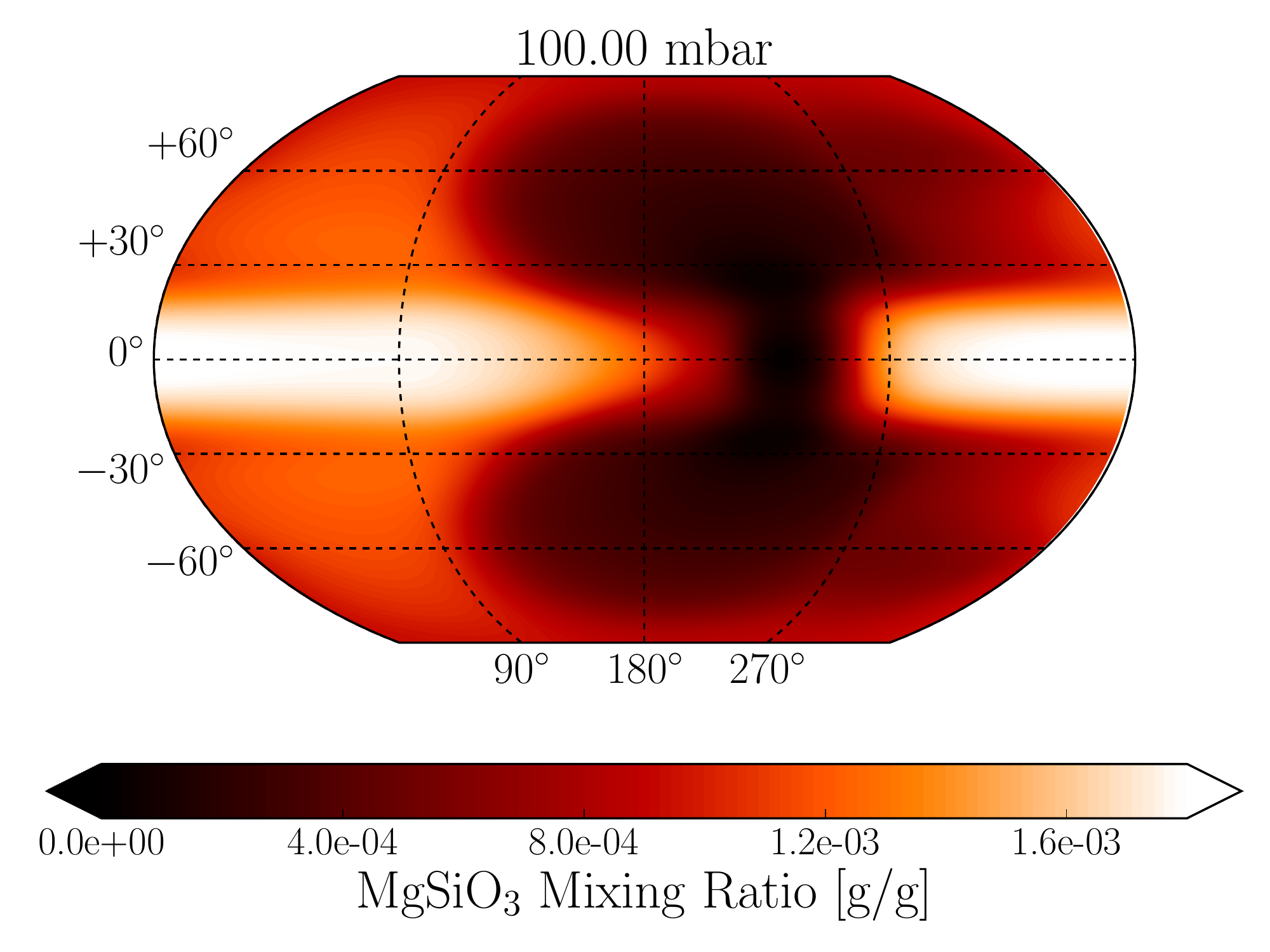}
\caption{}
\end{subfigure}\hspace*{\fill}
\begin{subfigure}{0.48\textwidth}
\includegraphics[scale = 0.44, angle = 0]{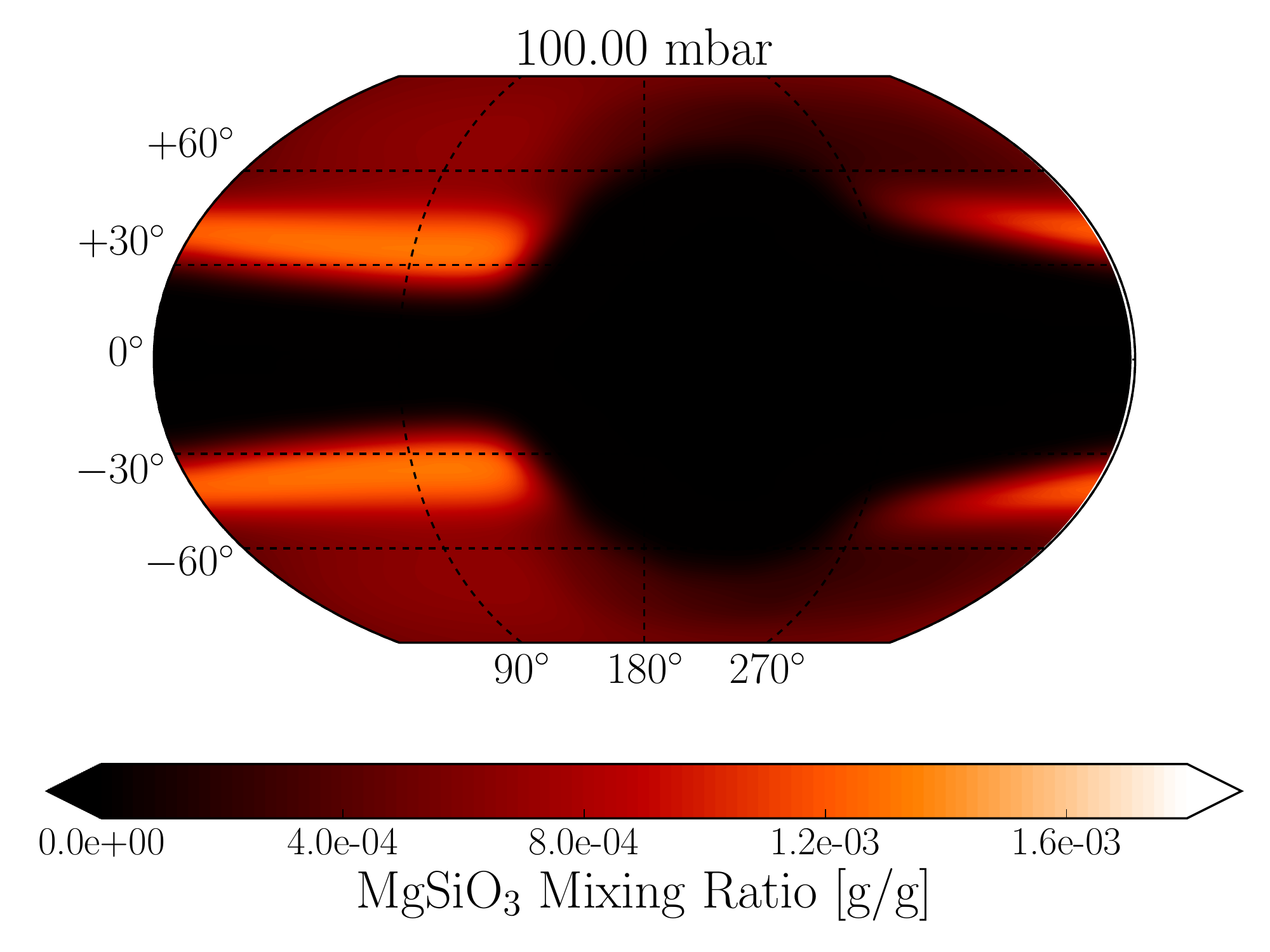}
\caption{}
\end{subfigure}
\medskip
\begin{subfigure}{0.48\textwidth}
\includegraphics[scale = 0.44, angle = 0]{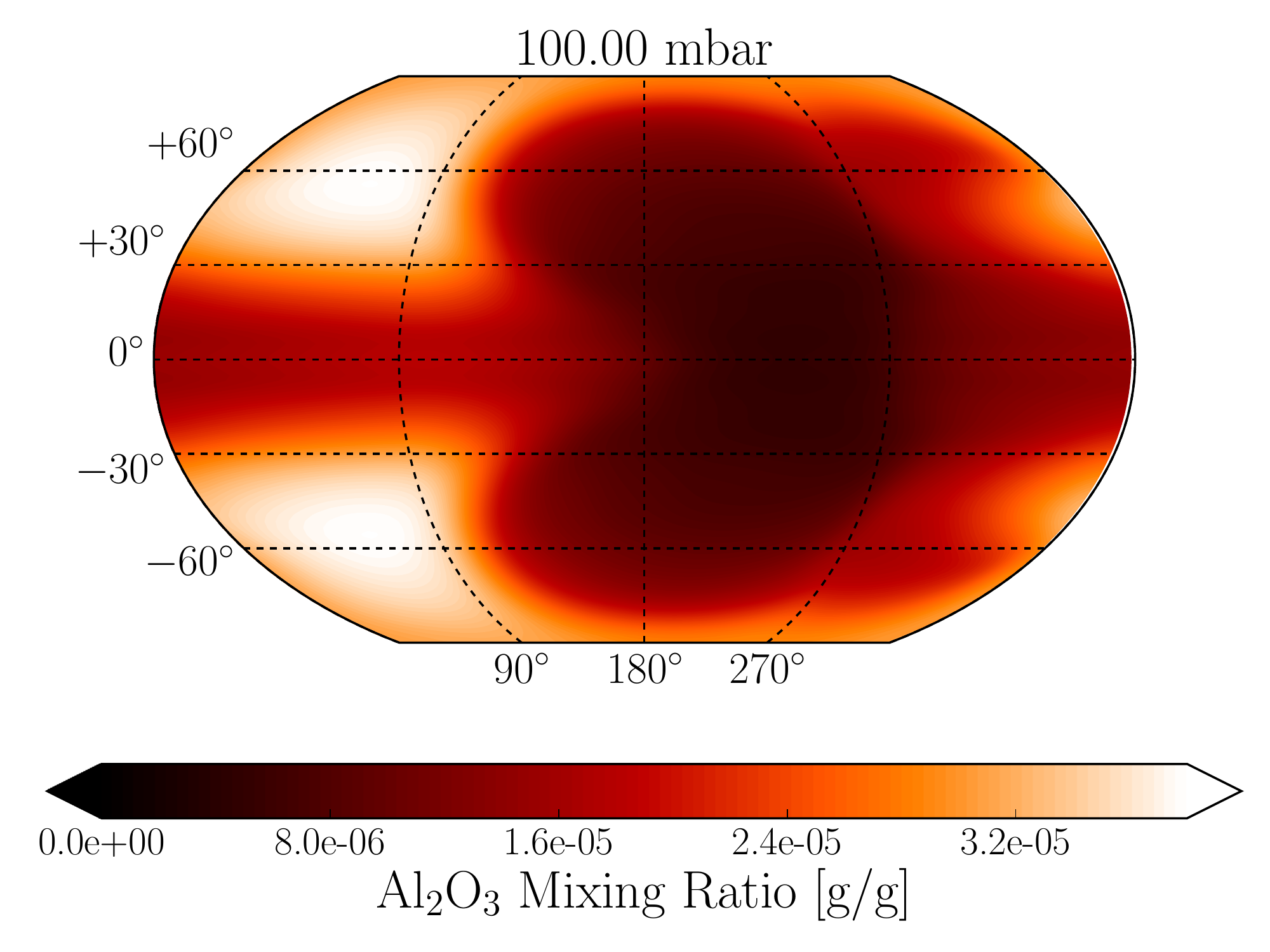}
\caption{}
\end{subfigure}\hspace*{\fill}
\begin{subfigure}{0.48\textwidth}
\includegraphics[scale = 0.44, angle = 0]{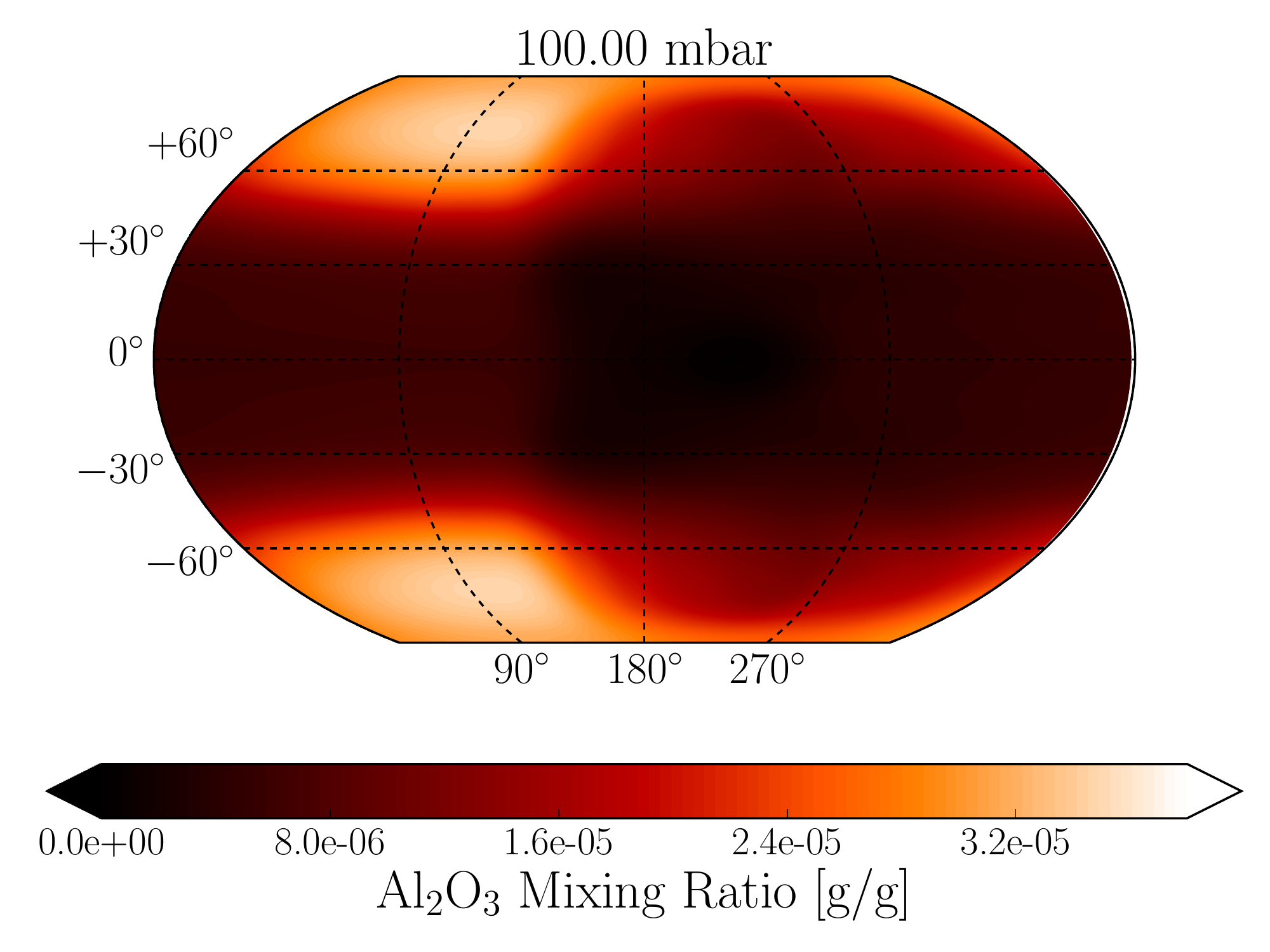}
\caption{}
\end{subfigure}
\caption{Cloud condensate mixing ratios of MnS, $\mso$ and $\alu$ (top, middle and bottom rows, respectively) for the hot deep interior and $\fsed$ = 0.1 simulation of HD~209458b. Data obtained during the initial diagnostic call at t = 0\,days with radiatively passive clouds (left) and final atmospheric state at t = 500\,days after the atmosphere has changed due to radiatively active clouds (right). The sub--stellar (dayside) point is at $\lambda$ = 180$^{\circ}$.}
\label{fig:hdi_f01_2d_cloud_latlon}
\end{figure*}

Figure \ref{fig:hdi_f01_2d_cloud_latlon} shows individual cloud condensate mixing ratios: MnS (top row), $\mso$ (middle row), and $\alu$ (bottom row) as a function of latitude and longitude at 100\,mbar. Figure \ref{fig:hdi_f01_2d_cloud_latlon} (top row) shows that the more volatile MnS is clearly depleted at the equator as well as across a large proportion of the dayside. The strongly irradiated dayside leads to a naturally higher temperature than the nightside. However, the well studied shift of maximum temperature from the sub--stellar point or `hotspot' shift, combined with the efficient advection of heat across the eastern terminator via high velocity winds in the super--rotating jet, cause an absence of MnS cloud over the first half of the nightside hemisphere. Within the jet itself, zonal wind speeds of up to 6\,kms$^{-1}$ help to sustain high temperatures for equatorial latitudes across the entire night side, high enough to restrict the condensation of MnS. 

The mid--latitude heating for P > 100 mbar, seen in Figure \ref{fig:hdi_f01_temp_1d} (right panel), for the anti--stellar point and west--limb results in an enhancement in the mixing ratio of MnS at 100 mbar across these regions, as the MnS base is forced to form at lower pressures due to strong the radiative heating beneath it. At 100\,mbar, the highly refractory $\alu$, (bottom row of Figure \ref{fig:hdi_f01_2d_cloud_latlon}), does not evaporate until around 2000\,K, a temperature that is not reached at any location within the simulated atmosphere at this pressure, leading to a global presence of $\alu$. The mixing ratio traces the local temperature similar to MnS, with notable decreases on the dayside and for equatorial latitudes.

There is, however, a hazard in over--interpreting these 2D mixing ratio maps, in that the abundance of cloud at a given pressure level is dependent upon, in addition to the eddy diffusion coefficient, the formation location of the cloud base. While the middle row of Figure \ref{fig:hdi_f01_2d_cloud_latlon} shows that $\mso$ conforms to the trend of decreasing abundance on the dayside, its increased abundance at the equator (compared to higher latitudes) opposes the trend seen for MnS and $\alu$. This effect can be explained by the sensitivity of a condensate to the deep atmosphere temperature and can be better understood using the mixing ratio profiles at t = 0\,days (dashed lines) shown in Figure \ref{fig:hdi_f01_mix_1d}. While for $\alu$ the cloud base forms at a similar pressure for both the equatorial and mid--latitudes, this is not true for $\mso$ where the base forms approximately 3000\,mbar deeper at $\phi$ = 45$^{\circ}$ than at the equator. The cloud abundance then falls off with altitude, towards lower pressures, at a similar rate for both latitudes which leads to a situation where for a given pressure level above the cloud base, the mixing ratio is higher at the equator. The formation location of the cloud base is, of course, dependent on the temperature structure. Considering the gas temperatures in Figure \ref{fig:hdi_f01_temp_1d}, it is evident that there is a large temperature gradient with latitude, particularly at pressures where the $\mso$ cloud base forms. Thus, we stress the importance of the deep atmosphere profile on the cloud structure at lower (and potentially, key observational) pressures. This effect also applies, albeit less clearly, for $\alu$. The extremely high condensation temperature of $\alu$ results in the cloud base forming at deeper pressures where the latitudinal temperature gradient reverses, allowing for a deeper base at lower latitudes. We explore the impact of the deep atmosphere temperature--pressure profile on the cloud structures in more detail in Section \ref{subsubsec:deep}.

The radiative properties of cloud particles are controlled by not only their chemical composition, but also their physical size. From Figure \ref{fig:hdi_f01_mix_1d} (third and fourth columns), particle sizes are shown to range from 80\,$\mu$m at around 10\,mbar to 10\,$\mu$m in the uppermost atmosphere. The effective radius appears to be far less sensitive than the mixing ratio to condensate type, latitude, longitude or even temperature--pressure profile. There exists a modest change in effective radius due to temperature changes caused by radiatively active cloud, with a small ($\sim$ 10\,$\mu$m) increase seen for most species.

Overall, considering the cloud abundance as the driving factor, the modification of the atmosphere's thermal state due to cloud radiative feedback drives a significant difference in the vertical cloud properties, and therefore retrieving the true cloud abundance is dependent upon solving for the cloud opacity feedback on the atmosphere's temperature--pressure structure.

\subsection{Cloud Properties}
\label{subsec:cloud_prop}

\subsubsection{Effect Of Sedimentation Efficiency, $\fsed$}
\label{subsubsec:sed}

Comparison of our simulations adopting `extended' $\fsed$ = 0.1 and `compact' $\fsed$ = 1.0 clouds allows us to explore the sensitivity of the cloud vertical structure and the atmosphere's final radiative state, to this `free' sedimentation parameter. Figure \ref{fig:all_cloud_1d}, introduced in Section \ref{subsubsec:gen_trend}, shows the total cloud mixing ratio as a function of pressure, for each of our simulations. At the sub--stellar point (red lines), the differences between the $\fsed$ = 0.1 and $\fsed$ = 1.0 simulations for the HDI case (top row of Figure \ref{fig:all_cloud_1d}) are significant. While the value of $\fsed$ determines the vertical extent of the cloud, it does not alter the location of the cloud base; this value is dependent on the local temperature and pressure only. Therefore, the total cloud mixing ratio for both the $\fsed$ = 0.1 and $\fsed$ = 1.0 cases, start at the point the highest temperature condensate forms, $\alu$, at 4 $\times$ 10$^4$\,mbar. The sedimentation parameter then strongly influences the behaviour of the condensate profile above the cloud base, toward lower atmospheric pressures. While the $\fsed$ = 0.1 case (top left, Figure \ref{fig:all_cloud_1d}) shows an almost linear decrease in the total cloud mixing ratios up to around 1\,mbar (until the Fe deck starts), the $\fsed$ = 1.0 case (top right, Figure \ref{fig:all_cloud_1d}) results in an almost `tri--deck' configuration, with three cloud mixing ratio maxima. The higher sedimentation efficiency effectively forces cloud into a more vertically compact structure, with lofted cloud above the base attenuated more quickly. This, therefore, results in an $\alu$ deck at the deepest pressure, a silicate deck in the mid--atmosphere and then a `volatile' (MnS and Cr) deck in the upper atmosphere.

Typically, the restriction of the cloud to higher pressures, deeper in the atmosphere, results in a weakened opacity for the upper, low pressure, atmosphere, and therefore reduced radiative impact, compared to the more vertically extended cloud, matching one of the conclusions of \cite{roman18}. As shown in the temperature--pressure profiles for each of our simulations in Figure \ref{fig:all_temp_1d}, the change in temperature when including cloud radiative feedback is lower for the $\fsed$ = 1.0 cases (right panels), than for the $\fsed$ = 0.1 cases (left panels).

For cloud that forms deep bases, such as the highly absorbing Fe and $\alu$, their presence in the upper atmosphere is reduced in the $\fsed$ = 1.0 simulations due to the enhanced, parameterised sedimentation. In some cases the majority of the cloud deck may lie below the photosphere. Indeed, even for the HDI atmosphere the $\alu$ base forms well below the photosphere levels shown in Figure \ref{fig:cf}. While this is shown to extend to low pressures in Figure \ref{fig:all_cloud_1d} (upper left) the increase sedimentation efficiency in Figure \ref{fig:all_cloud_1d} (upper right) shows a single cloud deck to be isolated below P = 10$^4$ mbar which is below the clear sky photosphere. The more reflective cloud species such as MnS and $\mso$ will also experience vertical compression such that their abundance is lowered in the very lowest pressure regions, but there is a still a significant contributed opacity from them, since their base forms at much higher altitudes and the cloud base pressure in unaffected by the $\fsed$ value and set only by the local PT profile.

For the majority of the atmosphere, at the equator, dayside heating occurs for both simulations i.e, $\fsed$ = 0.1 and $\fsed$ = 1.0. For pressures of 10 - 1000\,mbar, this temperature change is around 100\,K for our simulation with $\fsed$ = 1.0, 50\,K less than the $\fsed$ = 0.1 cases. In the upper atmosphere at P < 10\,mbar the low cloud opacity, mostly due to low abundances of MnS and Cr cloud, result in increases of less than 50\,K in temperature when including the radiative effect of clouds. However, even small changes in temperature can be enough to transition across the condensation curve and remove a particular cloud species. 

\subsubsection{Effect Of Deep Atmosphere Temperature}
\label{subsubsec:deep}

It has been suggested in a number of studies \cite[e.g. ][]{spiegel09,parmentier16,lines18a,powell18} that the temperature of the deep atmosphere can affect the distribution and hence observable signatures of cloud in hot Jupiter atmospheres. Similar to \citet{lines18a} we use two initial temperature--pressure profiles for HD~209458b. The SDI atmosphere exhibits a very different temperature structure at the highest pressures to the high entropy or HDI case. This can be seen via the temperature--pressure profiles in Figure \ref{fig:all_temp_1d} by comparing the form of the simulations at t = 0\,days (dashed) between the simulations; a temperature inversion occurs for the SDI simulation at around 6000\,mbar. The low temperature and high pressures (since increasing pressure allows for a higher condensation temperature) allow cloud to form down to the simulation inner boundary at 200\,bar, for the SDI simulations. \\ \\ \\ \\

Since, in the EddySed model, and therefore our simulations, cloud is effectively mixed upwards from the cloud base, the mixing ratio profiles in Figure \ref{fig:all_cloud_1d} shows that even for the low sedimentation efficiency ($\fsed$ = 0.1) simulation (left column), there is a much lower abundance in the mid and upper atmosphere for the SDI simulation compared to that of the HDI simulation. For some situations, there is an extraordinary change in cloud abundance; at the sub--stellar point, for the SDI $\fsed$ = 1.0 simulation (Figure \ref{fig:all_cloud_1d}, bottom right), there is a very vertically compact cloud deck (composed of $\alu$, Fe, $\mso$, $\msotwo$ and Cr) at the base of the simulated domain, with a small (both in pressure depth and by mass) deck of MnS in the very upper atmosphere (P > 1\,mbar). Since the abundance of the upper MnS deck is so low (< 10$^{-6}$\,g/g) and the remaining cloud exists at high pressures, deep in the atmosphere, there is no significant thermal change in the upper and mid atmosphere, and only a small increase in temperature at the inversion, when cloud radiative feedback is included.

The deep atmosphere temperature--pressure profile of hot Jupiters remains unconstrained; while we simulate an atmosphere with a physically motivated hotter deep atmosphere temperature, the solution could indeed vary over a wide range of temperature profiles. Since there is no significant change in the temperature--pressure profile between t = 0\,days and t = 500\,days for the SDI $\fsed$ = 1.0 simulation (Figure \ref{fig:all_temp_1d}, bottom right), we cannot clearly distinguish between an atmosphere (of any temperature profile) without cloud and one in which efficiently precipitating cloud has become cold--trapped \citep{parmentier16}.

\subsubsection{Modelling Approach}
\label{subsubsec:approach}

The atmospheric evolution and final thermal state differs between our microphysical simulations \citep[see][]{lines18a} and those in this work that implement a parameterised, phase-equilibrium model. In \cite{lines18a}, there is a global decrease in temperature which we attribute to the presence of high--opacity, highly--scattering, sub-micron mixed--composition silicate particles which are efficient at blocking the stellar insolation which, in a previously simulated cloud--free atmosphere, resulted in a warmer equilibrium state. In this work, parameterised vertical settling reduces the cloud abundance in the upper--most layers, allowing stellar flux to penetrate deeper into the atmosphere before being absorbed by gas--phase chemical species, such as H$_2$O and CH$_4$, or prior to radiative interaction with the cloud itself. Here, we briefly address the differences that arise due to our choice of cloud scheme.

In the EddySed model the physical size of cloud condensate particles is set via their effective particle radii, which are defined through the sedimentation parameter, $\fsed$, in addition to the applied lognormal distribution. Although dependent on both condensate--species and pressure, these homogeneously formed cloud particles typically extend from 10 - 80\,$\mu$m, making them significantly larger entities than the mixed--composition, kinetically formed particles in the microphysics--coupled model of \citet{lines18a}. It is worth noting that these particle sizes are obtained despite considering each cloud species independently; should they be able to undergo heterogeneous growth or coagulation as per the microphysics simulations of \cite{lee16} and \cite{lines18a}, then particles could potentially attain even larger radii (providing the upwards mixing is able to support them against precipitation). Larger particles can contribute a consistent opacity across a large range of wavelengths \citep{wakeford15} and not just focused at shorter wavelengths that can drive dominant scattering processes. This large contrast is important to address as for models that calculate the explicit particle settling velocity, the precipitation efficiency of the cloud will be strongly dependent on the particle size. Moreover, the physical size of the cloud particles will have a consequential effect on the radiative interactions, with larger particles introducing a reduced wavelength dependence to their opacity, and reducing the overall radiative cross-section per unit mass. The difference is caused by a combination of factors: the omission or inclusion of cloud condensation nuclei, the phase equilibrium or non--equilibrium approaches combined with the contrasting bottom--up or top--down model perspectives of EddySed and the microphysics model included in \citet{lines18a}, respectively. 

Exploring each of these components in turn, firstly, {\emph{EddySed}} assumes only homogeneous condensate particle formation and does not include the microphysics of seed particle nucleation. The added layer of complexity of seed particle nucleation which is included in \citet{lines18a} controls the number of potential condensation sites and thus can limit the overall particle sizes. Assuming constant condensation, a low or high rate of nucleation can moderate the particle sizes. For example, in \citet{lines18a} the low nucleation rates near the cloud base force the large available condensate mass onto a limited number of seeds, resulting in larger (micron sized) particles. In the upper atmosphere, high nucleation rates but a lower available condensate mass results in tiny sub--micron particles. Since physical size is a driving factor behind the outcome of a radiative interaction between stellar or thermal radiation and a given cloud particle, future modelling efforts should endeavour to focus on obtaining the most realistic particle size distribution possible. One potential solution for the microphysics approach is to implement a size bin scheme. \citet{powell18}, for example, use the CARMA (Community Aerosol and Radiation Model for Atmospheres) model which explicitly calculates particle sizes, handling the integration of each particle size bin independently. This numerically expensive procedure may be a necessary model component however, as their work shows that the size distribution of silicate grains in their simulations of a hot Jupiter atmosphere is bi--modal and partly irregular, making the post--application of an assumed size distribution a potentially poor approximation. However, it may be possible to use the size distribution predicted by these microphysics models, to inform parameterised schemes such as EddySed.

Secondly, EddySed assumes a phase--equilibrium combined with a bottom--up modelling approach, where the cloud base is formed at higher pressures first and, effectively, mixed upwards with a vertical extent dependent on the value of $\fsed$ and $K_{zz}$, as discussed previously. However, the microphysics model employed in \citet{lines18a} adopts a kinetic, non--equilibrium approach, meaning that cloud particles form over time, potentially never reaching a steady state as they advect through regions of the atmosphere with variations in local conditions. Additionally, seeds form and grains grow, initially, at the top of the atmosphere before precipitating down via gravitational settling. As found in \citet{lines18a}, tiny particles form in the upper--most atmosphere and subsequently struggle to efficiently settle down to the deeper layers; this inefficient precipitation or settling of cloud particles occurs despite \cite{lines18a} considering only the cloud's dynamical ascent via atmospheric circulation. Suspended cloud particles therefore exhibit poor growth via condensational growth due to being limited by the lack of available material. This situation is at contrast with the work presented here where the eddy diffusion parameter envelopes an unknown combination of both intra--cell advection and sub--grid vertical mixing, but does not consider the explicit vertical velocities from large scale flows. Should the true vertical velocities be incorporated in the approximation of the vertical mixing, then cloud particles may be lofted to lower pressures and potentially alter the strength of radiative feedback (as has been shown via the effect of sedimentation parameter in Section \ref{subsubsec:sed}). Incorporating the effects of the large scale circulation is not straightforward and, combined with the desire to match 1D approaches more closely, we reserve the the inclusion of this element for a future study.

The most challenging aspect of simulating atmospheres adopting a microphysical cloud model, with explicit sedimentation, are the computational resources required to evolve the atmosphere to an equilibrium between the vertical transport and sedimentation of the smallest cloud particles. However, in the case of EddySed the cloud base is formed at the highest pressure saturation point, before extending vertically up through the atmosphere. Since Eddysed implicitly assumes the vertical equilibrium of condensate material, particles can reach much larger sizes than those in microphysical simulations, as smaller particles in the upper atmosphere are assumed to have settled down to deeper atmospheric layers. Despite the intriguing differences in the simulated cloud structure between these two modelling approaches, no direct comparison between our simulation in this work and those of \citet{lines18a} should be made, as the latter does not consider the full suite of condensates included in this work. Performing a study which directly compares simplistic cloud formation schemes against microphysics models within the same surrounding, 3D, computational framework, is an important endeavour we aim to address in a future work.

\subsection{Synthetic Observations}
\label{subsec:synth_obs}

As discussed in Section \ref{subsec:model} our model can execute a second, higher resolution (500 bands), call to the radiative transfer scheme in order to provide high--resolution emission and transmission spectra. In this section we present synthetic observations derived from our simulations using this self--consistent numerical approach \citep[as employed in][]{lines18a,lines18b}, discussing transmission and emissions spectra, apparent albedo, as well as phase curves (in Sections \ref{subsubsec:trans}, \ref{subsubsec:emiss}, \ref{subsubsec:albedo} and \ref{subsubsec:phase}, respectively). We show that the presence of cloud can flatten the transmission spectrum, but each simulation produces identifiable atomic and molecular absorption features despite the presence of the cloud opacity. In the emission, the cloud opacity is shown to also weaken molecular absorption in the WFC3 bandpass. All our simulations are found to be consistent with measurements of the upper limit of the geometric albedo found by \cite{rowe08}. Finally, we find an excellent fit in terms of relative flux contrast, and a better fit than any previous 3D simulation of HD~209458b, to the 4.5 $\mu$m $Spitzer$ data from \cite{zellem14}.

\subsubsection{Transmission}
\label{subsubsec:trans}

For each model, the transmission spectrum is calculated by finding the total flux transmitted through the terminator and expressing it as a transit radius ratio. In our 3D transmission scheme \citep[see][for more details]{lines18b}, each column is treated independently and the calculation of the transmitted flux only includes the atmospheric properties from the columns on the night side of the planet limb. To offset the bias that can be introduced by an asymmetric opacity, such as the absence and presence of cloud across the day and night side of the terminator respectively, we perform an additional calculation of the transmitted flux whereby the properties are sampled from the dayside. The final spectrum then considers a simple mean of the fluxes obtained from each method, approximating the path of stellar photons which traverse atmospheric conditions from both day and night hemispheres. \citet{caldas19} provide an insight into the non-linear properties of the atmospheric limbs, and highlight some of the physical biases that should be considered when discussing transmission spectra.

\begin{figure*}
\includegraphics[scale = 0.4, angle = 0]{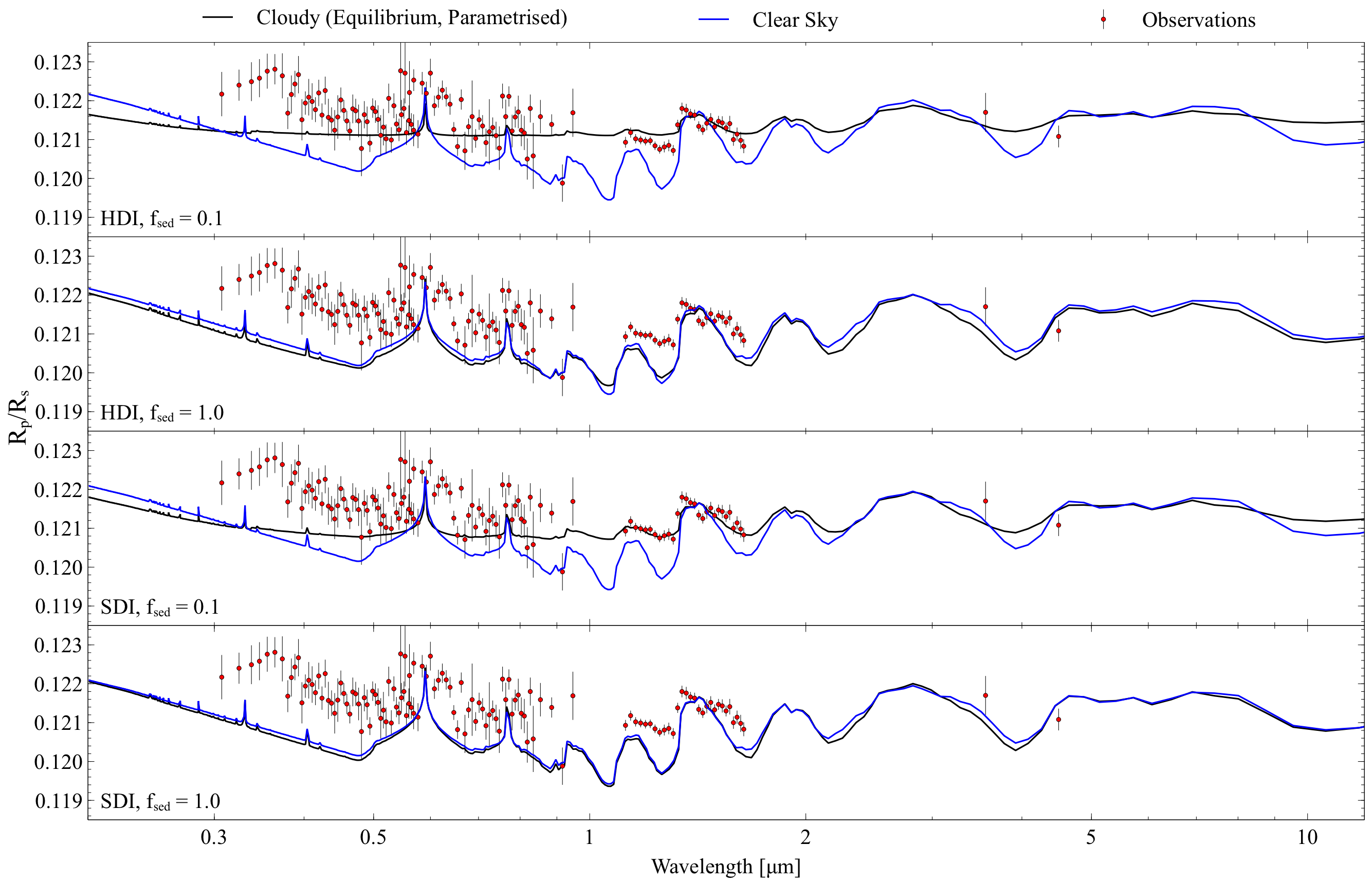}
\caption{Transmission spectra, sampled at t = 500\,days for all four of our simulations (black lines) with the clear sky (no cloud) spectrum at t = 0\,days (blue lines), and observations from \protect{\citet{sing08}} (red symbols) included. All spectra are normalised to the observations at $\lambda$ = 1.4\,$\mu$m due to the degeneracy of transmission spectra with the assumed planetary radius, and atmospheric extent.}
\label{fig:transmission}
\end{figure*}

Figure \ref{fig:transmission} presents transmission spectra (planetary to stellar radius ratio) from all four of our simulations comparing a `clear' sky (blue lines) and `cloudy' case (black lines), overlaid with the observations of HD~209458b from \cite{sing16}, where the simulated spectra are normalised to the observations at 1.4\,$\mu$m.  For the clear sky case the calculation is performed at the start of the simulation (t = 0\,days) where cloud is not present in the atmosphere, i.e. has not affected the thermal structure of the atmosphere via radiative feedback, nor contributed opacity to the flux calculation. The cloudy spectrum is calculated at  t = 500\,days where the cloud has both adjusted the thermal structure of the atmosphere, and contributes opacity to the derivation of the synthetic observables. When cloud is vertically extended ($\fsed$ = 0.1), Figure \ref{fig:transmission} shows both the HDI (upper two rows) and SDI (lower two rows) simulations show a significant flattening of the spectrum when cloud opacity is included. This is particularly true for the HDI case where the elevated cloud base coupled with the less efficient sedimentation results in a maximum cloud abundance (opacity) in the uppermost atmosphere. The condensate opacity competes with the gas phase opacity and leaves only weakened signatures from the alkali metals and water vapour. These simulated spectra do not show `super--Rayleigh' scattering, although this is to be expected due to the large effective particle radii which do not contribute to Rayleigh scattering in the optical or near--UV. Additionally, despite the high abundances of $\mso$ and $\msotwo$ condensates in our models, we do not see the silicate absorption found by \citet{lines18b} at 9 - 12\,$\mu$m. We attribute this to the increased opacity from iron and corundum clouds which act to smooth out any defined cloud--chemical signature from a single condensate species, and also due to the larger particle sizes found in this work, which contribute a less wavelength dependent opacity than smaller particles. The latter was tested by obtaining the transmission for an atmosphere that includes only silicate species (not shown). Overall, the cloud opacity appears to exert a strong influence across the 0.2 - 12\,$\mu$m wavelength region shown; the similar transit radius ratio across this range suggests grey cloud would be an acceptable approximation of our simulation results. 

Figure \ref{fig:all_cloud_1d} (bottom left panel) shows that despite the deep cloud base in the SDI simulations, a value of $\fsed$ = 0.1 still permits a significant cloud presence in the region of the atmosphere contributing to the transmission spectrum, due to the strong upward mixing of cloud condensate. However, this abundance and therefore opacity is reduced compared to the HDI case and results in a slightly less opaque atmosphere. The transmission profile is similar to that of the HDI atmosphere, but all the gaseous spectral features have an increased amplitude, over the HDI case, due to large amount of cold--trapped cloud below the photosphere. Efficiently precipitating cloud, described by our $\fsed$ = 1.0 simulations, only introduces subtle changes to the transmitted flux. This is in part due to the deep cloud introducing a much weaker opacity in the transmission region, as well as generating a much less pronounced temperature change from radiative feedback. While the deep cloud ($\fsed$ = 1.0) in the SDI case results in almost no difference compared to the clear sky atmosphere, for the HDI case the cloud opacity acts to block the flux windows at 1.05\,$\mu$m and 1.5\,$\mu$m which reduces the amplitude of the water vapour features in the near--IR. This indicates that even well settled, compact cloud could potentially be detectable in HST WFC3 observations. 

Figure \ref{fig:transmission} shows that our $\fsed$ = 0.1 cloudy atmosphere simulations, and in particular the SDI case, match the observations of \citet{sing16} most closely, despite our linear transit radius ratio with wavelength trend in the optical. Although the model spectra are normalised to the observations at 1.4\,$\mu$m, the cloud--induced flattening of the spectrum in the near--IR appears more consistent with the data. A wider size distribution that allows for more small--particle species to scatter in the optical while maintaining the grey profile in the near--IR may enhance the fit to the data. This reinforces the importance of studying further the potential cloud particle size distributions.

\subsubsection{Emission}
\label{subsubsec:emiss}

\begin{figure*}
\includegraphics[scale = 0.8, angle = 0]{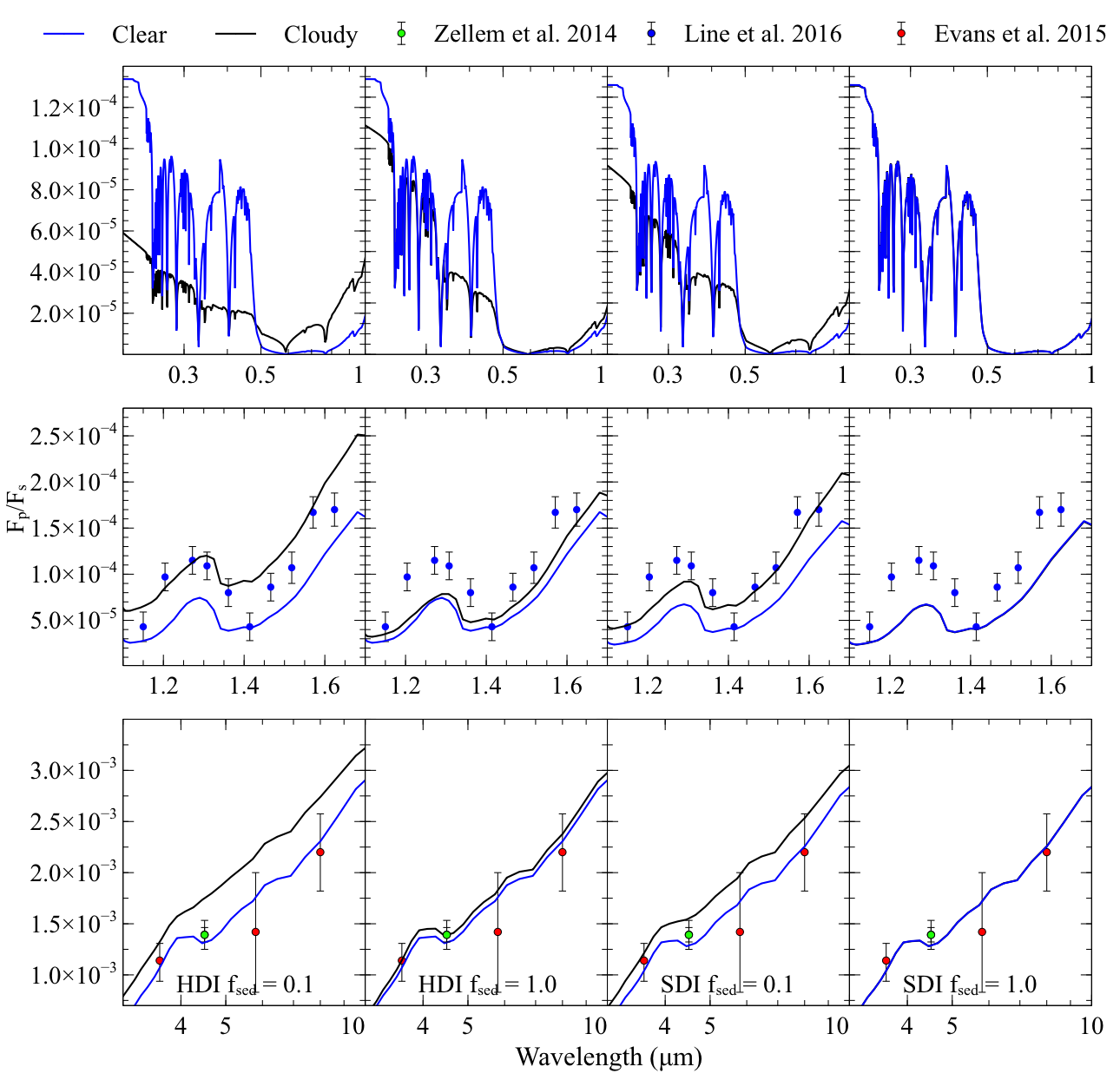}
\caption{Dayside `clear' sky at t = 0\,days with no cloud opacity (blue lines) and `cloudy' (black lines) emission at 0.2 - 1.0\,$\mu$m (top row), WFC3 G141 1.1 - 1.7\,$\mu$m (middle row) and 3.5 - 10\,$\mu$m (lower row), sampled at t = 500\,days, for all four simulations: both hot and standard deep interior profiles and $\fsed$ = 0.1 and 1.0. For the thermal emission, we include observations of the dayside emission from \protect{\citet{zellem14}} (green symbols) and \protect{\citet{evans15}} (red symbols), and for the WFC3 bandpass we include observations of the dayside emission from \protect{\citet{line16}} (blue symbols).}
\label{fig:emission}
\end{figure*}

Figure \ref{fig:emission} shows the dayside emission spectra for each of our four simulations, for the near--UV and optical (0.2 - 1\,$\mu$m, top row), near--IR WFC3 G141 (1.1 - 1.7\,$\mu$m, middle row) and IR (3.5 - 10\,$\mu$m, bottom row) wavelengths. Each plot considers the summation of the flux from the short--wave (reflected stellar component) and long--wave (thermal planetary emission) diagnostics from our radiative transfer scheme. As in Figure \ref{fig:transmission}, described in Section \ref{subsubsec:trans}, Figure \ref{fig:emission} includes both `clear' (blue lines) and `cloudy' spectra (black lines), derived at the start of the simulation (t = 0\,days) free of cloud contribution, and at the end of the simulation (t = 500\,days) including both the radiative impact on the atmospheric thermal structure and cloud opacity, respectively. For the HDI and $\fsed$ = 0.1 simulation (left column of Figure \ref{fig:emission}), since cloud radiative feedback leads to a large heating of the dayside, the emitted flux for $\lambda$ > 0.5\,$\mu$m is increased over that of the clear sky simulation. Additionally the cloud scattering opacity reflects photons before they can be absorbed by the broad Na and K wings, leading to a higher returned flux. Across the spectrum the signatures from gas absorption are less defined for the cloudy simulation as the cloud extinction is able to mask emission and reflection from gas lower in the atmosphere. Only in the near--UV for $\lambda$ < 0.5\,$\mu$m does the cloudy simulation emit less flux than its clear sky counterpart; cloud absorption (driven by the highly absorbing Fe and $\alu$ clouds that persist on the dayside - see equatorial and sub-stellar mixing ratio profiles in Figure \ref{fig:hdi_f01_mix_1d}) attenuates the reflected flux from H/He Rayleigh scattering. By considering this optical and near--UV cloudy flux from all our atmospheres, we note that vertically--equilibrated large--radii cloud condensates decks do not {\emph{necessarily}} contribute and enhanced flux at short wavelengths, providing the dayside condensates are composed of efficient absorbing species (in our case in particular, Fe). For example, in \citet{demory13} the simulated flux from 1D models in the Kepler bandpass is likely overestimated (providing iron and/or corundum clouds have not settled below the photosphere) due to their inclusion of only $\msotwo$ clouds in their atmosphere, a condensate that is strongly scattering. Since we have efficiently absorbing and high mass cloud condensates such as Fe and $\alu$ in the upper atmosphere, absorption rather than scattering, is generally the dominant factor. 

In the HST WFC3 bandpass, shown for the middle row of Figure \ref{fig:emission}, the flux from an atmosphere with radiatively active cloud can reach three times the emitted flux of the cloud--free case, a combination of both the enhanced thermal emission due to the atmospheric temperature increase and also the reflective component of clouds. With the exception of the SDI and $\fsed$ = 1.0 simulation, where our cloudy simulation shows no difference over the cloud--free emission, the opacity contribution of the cloud particles also diminishes the amplitude of the 1.4\,$\mu$m water feature (as seen for the case of transmission, Figure \ref{fig:transmission}); a spectrum with shallower absorption features across this wavelength range becomes a good indication of the presence of cloud. This is true even for our HDI simulation with well--settled and compact ($\fsed$ = 1.0) cloud (Figure \ref{fig:emission}, second column). \cite{line16} measure the dayside emission of HD~209458b using the HST WFC3 instrument, and we include their data (blue symbols) for comparison. With the exception of the data taken at $\lambda$ = 1.15 $\mu$m and $\lambda$ = 1.41 $\mu$m, we find a good match to the observations with our extended cloud ($\fsed$ = 0.1) simulations, both in absolute flux ratio but also with the amplitude of the H$_2$O feature. Generally, the evidence for dayside clouds is best related to the overall increase in flux within the WFC3 bandpass, rather than any discernible change to the spectral shape.

The IRAC data from \cite{evans15} (red symbols) and \cite{zellem14} (green symbol) are overlaid for the thermal emission Figure \ref{fig:emission} (bottom row) for all simulations. While the synthetic emission from our clear sky atmosphere is generally a closer match to the observations than the cloudy simulations, we do find a good match to all secondary eclipse datasets (IRAC and Spitzer) with our SDI and $\fsed$ = 0.1 simulation. \cite{zellem14} find a similar match to Spitzer data, using GCM simulations of \cite{showman09} that show a cloud--free atmosphere is a good match to the observed thermal emission. This does not rule out the presence of clouds in HD~209458b however as cloud is required to match the IRAC data, which is poorly modelled by our clear--skies GCM. This results indicates that the opacity difference between WFC3 and 4.5 microns is not large enough compared to the temperature gradient predicted from our simulations.

For the SDI simulations (third and fourth columns of Figure \ref{fig:emission}), the influence of the cloud on the emission for the $\fsed$ = 0.1 case is much stronger than for the $\fsed$ = 1.0. Essentially, the choice of deep atmosphere temperature--pressure profile, i.e. HDI or SDI, becomes less relevant in detecting cloud in emission providing the cloud is vertically extended (compare columns one and three of Figure \ref{fig:emission}). For the SDI $\fsed$ = 1.0 simulation (fourth column of Figure \ref{fig:emission}) there is almost no change in the emission between the clear and cloudy cases, alongside the small changes in transmission discussed in Section \ref{subsubsec:trans}. The sensitivity of the emission on the vertical positioning of the cloud is also found and discussed in \cite{roman18}.

\begin{figure}
\includegraphics[scale = 0.57, angle = 0]{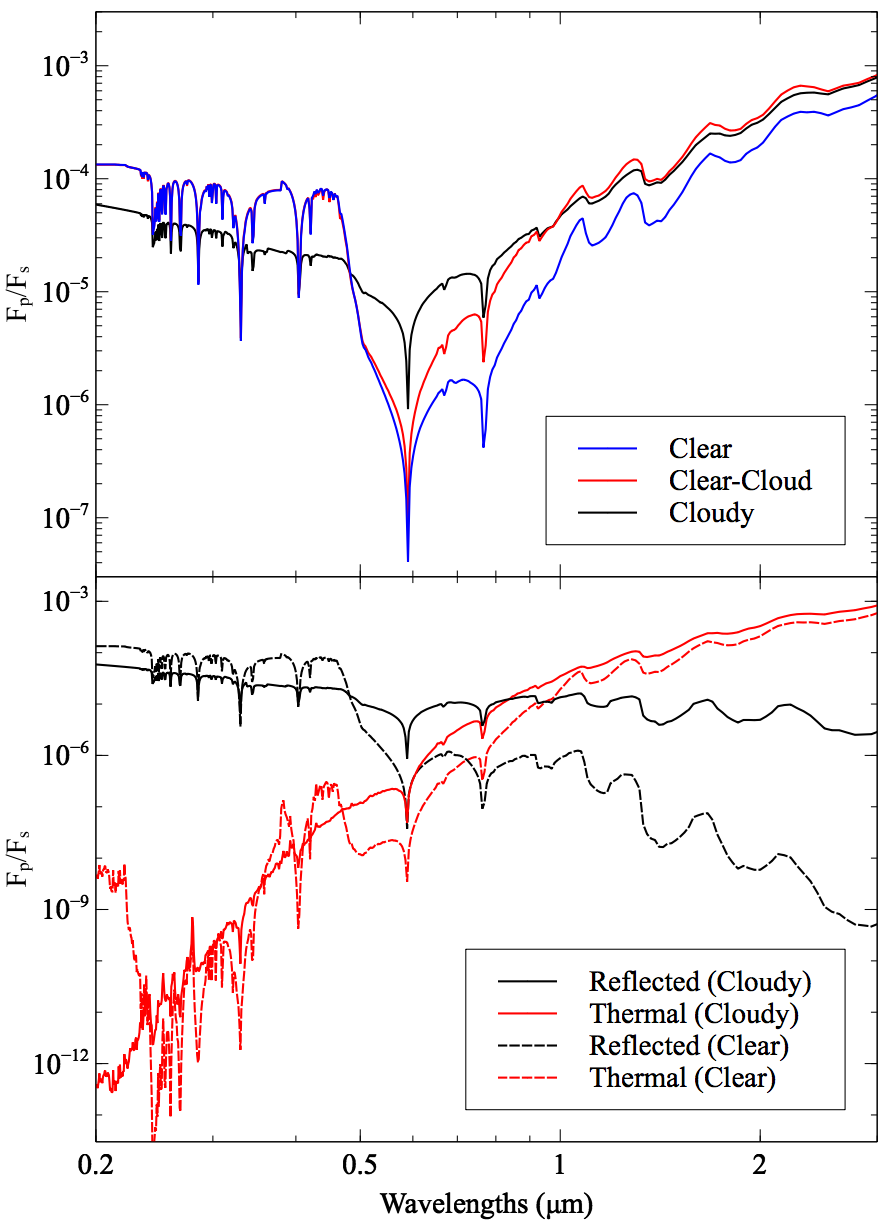}
\caption{Upper Panel: Dayside `clear' (blue line), `clear--cloud' (red line) and `cloudy' (black line) emission between 0.2 - 3.0\,$\mu$m for HDI and $\fsed$ = 0.1. The clear emission is from the simulation at 0\,days, without cloud opacity, the `clear--cloud' is the emission after 500\,days (when cloud radiative feedback has altered the temperature--pressure profile) but cloud opacity is not included in the flux calculation, and the `cloudy' emission is from the simulation at 500\,days but also includes the cloud opacity flux calculations. Lower Panel: reflected (black lines) and thermal (red lines) components of the emitted flux for the HDI and $\fsed$ = 0.1 (solid lines) and clear skies (dashed lines) simulations.}
\label{fig:emission_thermal}
\end{figure}

Figure \ref{fig:emission_thermal} (upper) presents the emission from the HDI and $\fsed$ = 0.1 simulations, isolating the effect of both the clouds themselves and the change in the temperature--pressure profile, and thereby thermal emission, caused by the radiative impact of the clouds. The emission is shown for three cases. Figure \ref{fig:emission_thermal} (lower) shows the individual reflected (black lines) and thermal (red lines) components of the `emitted' flux for the DI and $\fsed$ = 0.1 (solid lines) and clear skies (dashed lines) simulations. Firstly, the clear (blue line) and cloudy (black line) spectra, as previously defined in Section \ref{subsubsec:trans}, are shown i.e. the emissions at t = 0\,days without any cloud contribution, and that at t = 500\,days including the cloud impact on the temperature structure and cloud opacity. In addition, the emission at 500\,days (when the thermal profile has adjusted to the radiative effect of the clouds) but without the cloud opacity, termed the `clear--cloud' spectrum (red line) is also shown. In Figure \ref{fig:emission}, for $\lambda$ < 0.6\,$\mu$m (upper row) the clear--cloud spectrum closely matches the clear skies spectrum indicating that the temperature increase on the dayside itself, caused by the radiative impact of the clouds, does not affect the `emission' profile in the near--UV. This is to be expected as the thermal component of the emission at such short wavelengths is negligible. For $\lambda$ > 1\,$\mu$m the clear--cloud spectrum now switches to match, roughly, the cloudy emission. Small changes arise due to the active cloud opacity (a process which clearly weakens the absorption features in the near--IR and IR), but at these wavelengths the dominant factor in the change in the emission between the clear and cloudy simulation is the dayside temperature increase from radiatively--active, heating clouds. In the optical there is an intermediate situation in which the clear--cloud emission lies somewhere between the clear and cloudy cases, and is shown by the crossover of both clear and cloudy spectra at $\sim$ 0.5\,$\mu$m in Figure \ref{fig:emission_thermal}. Here the thermal component is important, albeit weaker than at longer wavelengths, but the cloud opacity itself also drives the increased flux. The individual contributions from the reflected and thermal light in the lower panel of Figure \ref{fig:emission_thermal} also assist in showing the transition from the dominance of reflected light for $\lambda$ < 0.8\,$\mu$m and thermal for wavelengths greater than 1\,$\mu$m.

\subsubsection{Albedo}
\label{subsubsec:albedo}

\begin{figure}
\includegraphics[scale = 0.57, angle = 0]{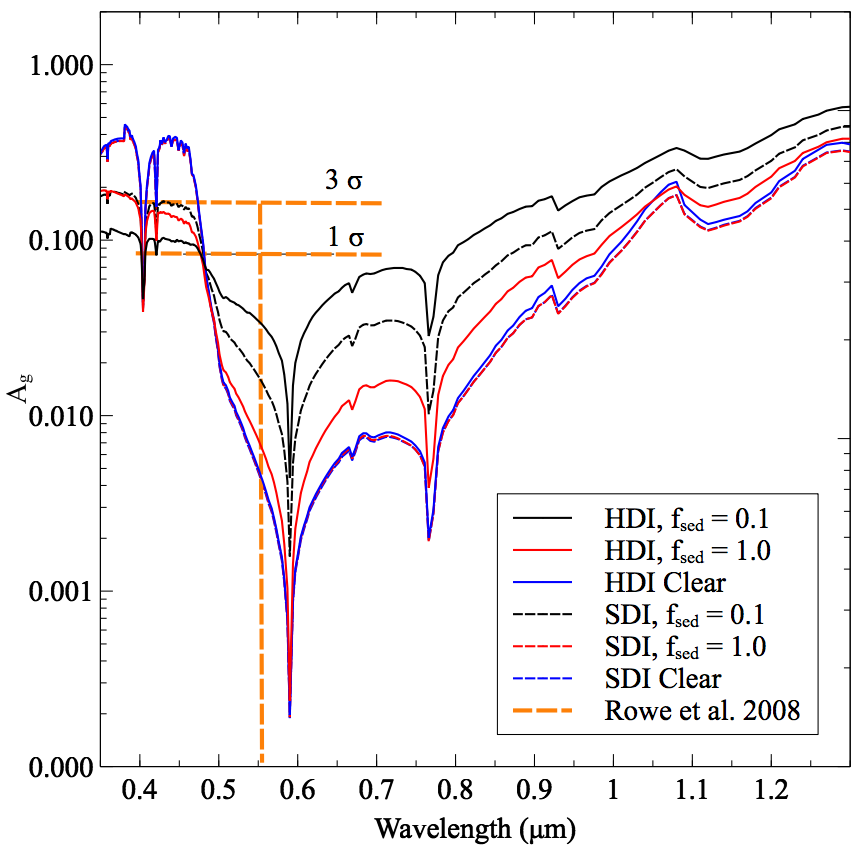}
\caption{Apparent albedo, $Ag$, between 0.35 - 1.3\,$\mu$m for HDI and $\fsed$ = 0.1 (solid black), HDI and $\fsed$ = 1.0 (solid red), Clear HDI (solid blue), SDI and $\fsed$ = 0.1 (dashed black), SDI and $\fsed$ = 1.0 (dashed red) and Clear SDI (dashed blue). Albedo data from \protect{\citet{rowe08}} are shown, with the 1$\sigma$ upper limit of 0.08 and the 3$\sigma$ upper limit of 0.17 as the lower and upper horizontal dashed orange lines respectively.}
\label{fig:albedo}
\end{figure}

Figure \ref{fig:albedo} presents the synthetic apparent albedos, ($Ag$), derived for clear sky (blue) and cloudy sky ($\fsed$ = 0.1, black and $\fsed$ = 1.0, red) cases for both the HDI (solid) and SDI simulations (dashed). Using {\textit{MOST}} (Microvariability and Oscillation of Stars), \cite{rowe08} obtain constraints on the upper limit of the albedo, and their 3$\sigma$ and 1$\sigma$ values are overlaid in Figure \ref{fig:albedo} (dashed orange). We convolved our synthetic fluxes with the {\textit{MOST}} transmission response function and obtain band--averaged (400 - 700 nm) albedos, $A_{g}$($\bar{\lambda}$), for each of our simulations. For both our SDI and HDI clear sky simulations we find $A_{g}$($\bar{\lambda}$) = 0.081, matching the observed 1$\sigma$ upper limit. Whereas, for both the cloudy HDI and SDI simulations with $\fsed$ = 0.1 we find $A_{g}$($\bar{\lambda}$) = 0.06. The lowest albedo is given by the HDI $\fsed$ = 1.0 simulation at $A_{g}$($\bar{\lambda}$) = 0.04 while the highest albedo for a cloudy simulation is for the SDI and $\fsed$ = 1.0 case ($A_{g}$($\bar{\lambda}$) = 0.08) since the deep compact cloud has little influence on the flux, as has been shown for both the transmission and emission. Despite the presence of clouds, our synthetic albedos are consistent with the upper limits from \cite{rowe08}. Although reflective clouds may enhance the geometric albedo through back--scattered light, the high atmospheric temperatures in our simulations (increased due to our radiatively active clouds) which act to `thin out' or remove reflective MnS and silicate clouds on the dayside, mean that the majority of our dayside flux is  moderated by highly--absorbing Fe and $\alu$ clouds (see equatorial, sub-stellar condensate mixing ratios in Figure \ref{fig:hdi_f01_mix_1d}). We therefore caution against assuming a clear sky atmosphere when fitting to measurements of low albedo, since an alternative explanation can include thick decks of radiatively active (and hence important to the solution of the TP profile), large particle and poorly reflective Fe and $\alu$ cloud.

\subsubsection{Phase Curves}
\label{subsubsec:phase}

\begin{figure*}
\includegraphics[scale = 0.25, angle = 0]{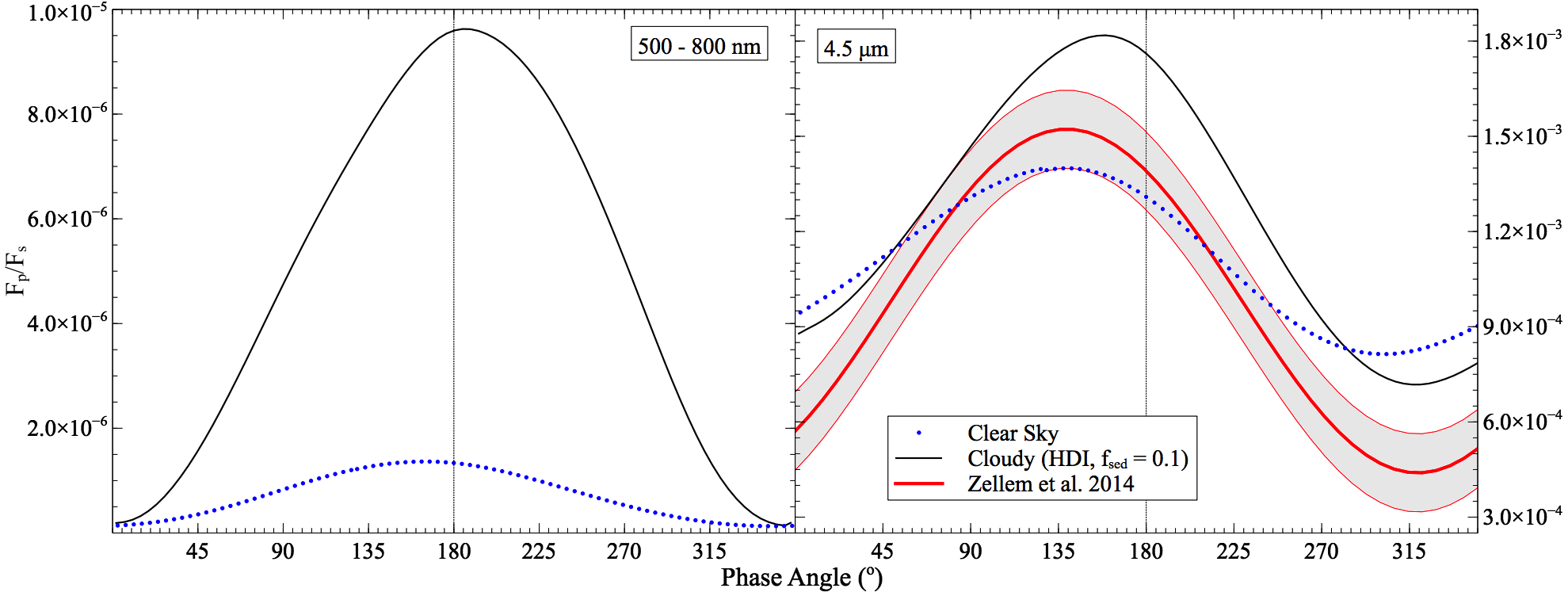}
\caption{`Clear' sky, cloud--free, spectrum at t = 0 days and omitting cloud opacity (dotted blue line) and the cloudy, t = 500\,days (black line) phase curve at 500 - 800\,nm (left) and 4.5\,$\mu$m (right), sampled at t = 500\,days, for the HDI and $\fsed$ = 0.1 simulation. Observations from \protect\citet{zellem14} (red line) with 1$\sigma$ error (shaded) are included for the thermal phase curve (right panel).}
\label{fig:phase_zellem}
\end{figure*}

In Figure \ref{fig:phase_zellem} we present the synthetic full orbit phase curve of our HDI, $\fsed$ = 0.1 simulation at $\lambda$ = 500 - 800\,nm (left panel) and $\lambda$ = 4.5\,$\mu$m (right panel), at 0 \,days without cloud opacity (blue dotted line) and 500\,days (red) including cloud opacity, with the \citet{zellem14} phase data from the Spitzer 4.5\,$\mu$m channel (red line and shaded region denoting 1$\sigma$ uncertainty) for the thermal phase curve overlaid (right panel). For the optical phase curve (left panel, Figure \ref{fig:phase_zellem}) it is well known that cloudy hot Jupiters can exhibit a westward shift \citep[e.g.][]{esteves13, shporer15,parmentier18}, potentially due to the presence of scattering, high--albedo cloud on the cooler western limb. In the optical, for 500 - 800 nm, we find a modest shift in the phase maxima, westwards of the sub--stellar point. Additionally, due to both the temperature increase from cloud radiative feedback and the reflective component of the cloud between these wavelengths, the cloudy atmosphere flux in this bandpass is around an order of magnitude larger than that obtained for the clear skies case. At the equilibrium temperature of HD~209458b, \citet{parmentier16} show that MnS is a likely candidate for asymmetric cloud coverage due to high dayside temperatures, and our simulations return this enhanced abundance of MnS, shown via the mixing ratio maps in Figure \ref{fig:hdi_f01_2d_cloud_latlon} (upper row). However, the magnitude of this offset is likely reduced by the rise in dayside temperature due to our cloud radiative heating; the extreme dayside temperatures force the strongly non--homogeneous MnS almost entirely out of the dayside, leaving behind high temperature condensates that are able to sustain themselves across the full extent of the dayside hemisphere, leading to less of an asymmetric distribution of cloud, albedo and hence optical flux.

For the thermal flux component (right panel, Figure \ref{fig:phase_zellem}), cloud--free GCM simulations of HD~209458b have been shown to overpredict the nightside flux \citep[e.g.,][]{zellem14,amundsen16,showman09}. Indeed, the flux from our cloud--free simulation, or clear sky, matches closely the observed dayside flux (within the 1$\sigma$ confidence), but significantly overestimates the nightside flux. The emission from the same simulation including the cloud opacity, i.e. cloudy, reveals an improved match to the nightside flux and a better match in absolute flux contrast, matching the observed phase amplitude almost exactly. However, the cloudy model now overestimates the dayside flux, and continues to overestimate the nightside. It is tempting to attribute the enhanced dayside emission due to the large temperature increase from cloud radiative heating, but the dayside temperatures obtained in our cloudy simulations are required in order to match the IRAC observations. As mentioned in section \ref{subsubsec:emiss}, this situation can be attributed to an opacity difference between the WFC3 band and 4.5 microns that is not large enough compared to the temperature gradient. 

Since the radiative properties of the cloud, and hence contributing opacity, are controlled by their chemical composition and physical particle or droplet size, there are a number of ways this situation may arise. Using a non--equilibrium cloud chemistry approach, for example, may help with the preferential settling of condensates: if iron particles formed quickly with large effective radii (due to the large initial abundance), they may well settle out of the atmosphere, resulting in a weakened opacity. We may be overestimating the particle radii with the applied size distribution; the considerably smaller particles in \citet{lines18a} can cool the atmosphere down globally, but by retaining condensates that have a preference for absorbing, it may be possible to maintain the overall zonal flux contrast. Finally, since in this work we approximate the vertical mixing with $K_{zz}$, smaller and more reflective particles may not reach altitudes as high as if we included the true vertical velocities found in our simulations. The complexities of including both sub--grid mixing and large scale vertical flows is something we leave to future work.

One of the key findings of 3D GCM simulations of hot Jupiters is the eastwards shift of the peak emitted thermal flux due to the shift of the `hot--spot'. In our simulations, Figure \ref{fig:phase_zellem} (right panel) shows the presence of clouds causes this eastwards offset to reduce by approximately 15$^{\circ}$ compared to the clear skies simulation. Temperature maps (not shown) indicate that although the peak temperature (location of the hot--spot) remains at a similar longitude when including cloud radiative feedback, the longitudinal temperature gradient at the equator is reduced due to the efficient heat redistribution from the equatorial jet. This shifts the overall emitted flux peak back towards the sub--stellar point. The ability for clouds, even those confined entirely to the nightside hemisphere, to decrease the offset of the phase curve maxima, is discussed in \cite{parmentier18}.


\section{Conclusions}
\label{sec:conclusions}
We first list our main conclusions, before expanding where required:
\begin{enumerate}
\item Including radiative feedback (scattering and absorption) from cloud condensates significantly alters the thermal structure of our simulations of HD~209458b, in turn, altering the cloud properties themselves. 

\item Cloud properties and therefore the atmospheric temperature structure once radiative feedback from the clouds is included, are strongly affected by the both the choice of the sedimentation efficiency ($\fsed$) and the choice of the initial deep atmosphere temperature--pressure profile (HDI/SDI). 

\item The radiative balance and final thermo--chemical structure of our simulated atmosphere with radiatively active cloud is markedly different to our kinetic, microphysics--coupled simulations presented in \citet{lines18a}.

\item Compared to our cloud--free simulations, we find an improved match to observations, in both the transmission and the near-IR (WFC3) and IR (IRAC) emission. In particular, we show good agreement with the dayside thermal emission of HD~209458b in the HST WFC3 wavelengths.

\item The apparent albedos from our cloudy simulations are consistent with the 1$\sigma$ upper limit obtained by \cite{rowe08} for HD~209458b.

\item We obtain an improved match to the thermal full orbit phase curve observations in the 4.5\,$\mu$m Spitzer channel, over previous 3D cloudy and cloud--free atmosphere simulations, but still overestimate the day and nightside emission. We find a modest westwards shift in the optical due to the presence of reflective, MnS cloud at the west--limb.

\end{enumerate}

Our simulations show a significant temperature difference between an atmosphere with passive (or post--processed) cloud and one which includes the effect of cloud scattering and absorption. Our optically thick condensates drive a global heating and strong day--night temperature contrast. The equilibrated temperature--pressure solution leads to the formation of cloud bases at low pressure, contributing a larger cloud abundance from refractory condensates, and complete or partial removal of volatile species, in the upper atmosphere.

The presence of cloud enables efficient atmospheric heating, with strong absorption from condensate particles at the shortest wavelengths. We also find that despite clouds raising the thermal photosphere, the planetary equilibrium temperature increases as well, since large positive temperature changes at the new photosphere level exceed the temperatures at the original (clear sky) photosphere level.

The importance of cloud radiative feedback depends on the abundance of cloud in the upper and mid atmosphere, where cloud particles are able to interact with the stellar insolation. The sedimentation efficiency and choice of initial deep atmosphere temperature profile strongly influence the vertical distribution of the cloud. An increased sedimentation efficiency leads to more compact cloud which concentrates the cloud abundance (and hence opacity) at higher pressures and results in a weak change to the atmospheric thermal state. The temperature of the deep atmospheres can moderate the cloud abundance in the upper atmosphere also, by changing the position of the cloud base; a hotter interior drives the cloud base to lower pressures where mixing then distributes a larger abundance of cloud to the upper atmosphere, driving a more significant change in the atmosphere's thermal state due to radiative clouds. 

While no true comparison can be made between our phase--equilibrium, parameterised cloud treatment in this work, and the kinetic, microphysical cloud in \citet{lines18a} we acknowledge that the choice of cloud model can result in a vastly different cloud structure and hence final equilibrium state of the atmosphere. In our microphysics--coupled model, we include fewer condensates (particularly the absorbing iron and corundum), obtain significantly smaller (sub-micron) cloud particles than the deci--micron particles in this work and also see a large number of suspended cloud particles in the uppermost atmosphere. We again caution that since the vertical mixing in this work is parameterised only through the value of $K_{zz}$, we may underestimate the abundance of smaller particles in the upper atmosphere due to the omission of the true vertical velocities; the vertical dynamics of cloud particles will be investigated in a future work. This difference in both condensate composition and particle size can drive either cooling through dominant scattering, or heating through absorption of the irradiation. Importantly, since one of the main controlling elements of the radiative balance is the cloud particle radius, motivating study of the potential size distributions of the cloud particles using model such as the size bin--scheme CARMA model. 

In terms of the synthetic observations derived from our simulations, each of our four simulations produce a transmission spectrum with identifiable absorption features. Even for the HDI and $\fsed$ = 0.1 case, which has the largest cloud opacity in the transmission region, water vapour and the prominent sodium and potassium alkali signatures are well defined. Vertically extended cloud at $\fsed$ = 0.1 greatly flattens the spectrum in comparison to the compact $\fsed$ = 1.0 case, reducing the amplitude of all absorption features, and also results in a weakened Rayleigh scattering slope at near--UV and optical wavelengths.

Additionally, in emission the cloud condensate opacity masks the 1.4\,$\mu$m water vapour feature, and we find a good match to HST WFC3 observations by \cite{line16} from our extended cloud ($\fsed$ = 0.1) simulations, and also to the \cite{evans15} and \cite{zellem14} IRAC measurements our SDI and $\fsed$ = 0.1 simulation. The presence of cloud enhances the flux in the infra--red from radiative heating but mutes the near--UV flux for $\lambda$ $<$ 0.5\,$\mu$m due to its absorption opacity. Cloud opacity, even for compact or deep forming cloud, reduces the amplitude of the 1.4\,$\mu$m water vapour absorption feature, making the identification of clouds in the WFC3 bandpass potentially possible. Only the SDI and $\fsed$ = 1.0 case (deep forming and compact clouds) sees no change in the emission spectrum. Additionally, the thermal emission is more strongly affected by the temperature increase due to cloud radiative feedback than the cloud opacity itself. By analysis of the clear--cloud component (radiatively adjusted atmosphere at 500\, days, but without the cloud opacity included in the flux calculations) of the near--IR and IR emission (see Figure \ref{fig:emission_thermal}), the flux change is most strongly altered by the thermal state of the atmosphere, due to the cloud radiative feedback, and not the cloud opacity itself. However, the cloud opacity itself is important in muting the emission features.

We show good agreement with the 1$\sigma$ upper limit on the geometric albedo found by \cite{rowe08}. Cloud absorption across the {\textit{MOST}} bandpass reduces the reflected flux from H/He Rayleigh scattering and reduces the apparent albedo from $Ag$ = 0.08 for our clear sky atmosphere to $Ag$ =0.04 for our SDI and $\fsed$ = 0.1 simulation. High dayside temperatures, driven by cloud radiative feedback, inhibit the formation of reflective silicate clouds, and the remaining Fe and $\alu$ cloud decks can efficiently reduce the albedo.

In the thermal emission, the temperature contrast that arises from cloud radiative feedback results in an improved match (via the absolute flux contrast) to observations in the 4.5\,$\mu$m {\emph{Spitzer}} channel, over our previous cloud, and cloud--free simulations \citep{amundsen16,lines18a}. Additionally, the zonally extended hot--spot that is driven by dayside cloud absorption, alongside a cooler, cloud--friendly mid--latitude atmosphere that can sustain cloud at the eastern limb, causes a reduction in the eastwards shift in the peak emission. This means that the presence of cloud can reduce the eastwards shift in the thermal phase curve. For the cloudiest atmosphere (HDI and $\fsed$ = 0.1) we see, in the optical, more than a three times increase in the flux amplitude over our clear sky simulation; such a large contrast is potentially detectable. The optical phase curve shows a limited shift of the phase curve maxima west of the sub--stellar point, and a significant shift overall from the clear sky atmosphere, confirming the ability for axisymmetric cloud to generate a westward offset across optical wavelengths. 

Finally, if cloud forms deep in the atmosphere (as per SDI), and is compact due to efficient settling ($\fsed$ = 1.0) then our cloudy and cloud--free simulations can not be easily differentiated, via analysis of the emission, transmission and phase curves. This holds even when considering the radiative effects of deep forming and compact clouds on the atmosphere's thermal structure. All other combinations of cloud (deep and extended, shallow and compact) make a significant impact on synthetic observations we have derived, including deep, cold--trapped cloud (SDI) providing it is vertically extended.

To summarise, we have shown via 3D radiative--hydrodynamic simulations, including a parameterised cloud formation scheme with phase--equilibrium chemistry, that many cloud condensates are potentially present in the atmospheres of HD~209458b and indeed likely in many hot--Jupiter atmospheres; \cite{parmentier16} find using an equilibrium cloud code that condensates are likely to exist over a wide range of planetary equilibrium temperatures, with the precise composition sensitive to the atmosphere temperature. Moreover, these condensates can play a significant role in the overall radiative balance of the atmosphere. Unlike the cooling atmosphere found in \cite{lines18a} with microphysical clouds, the condensate composition (chemical and physical) and distribution predicted using the EddySed model typically raises both the dayside and nightside temperature, but larger dayside heating rates, compared to the night hemisphere, leads to a increased day--night contrast over a clear sky atmosphere. With $\alu$ included, we find a similar behaviour (warming atmosphere) to \cite{roman18} who find their double--grey radiative clouds have a significant feedback on the atmospheric state of Kepler-7b.

Their presence has a significant and potentially detectable effect on the transmission and dayside emission spectra, as well as phase curve observations. However, the EddySed model may well overestimate both the efficiency of settling and effective radii of condensate particles (potentially driven by the exclusion of the true vertical velocities from large scale atmospheric flows), whereas our implementation of the \citet{helling13} model in \citet{lines18a} likely underestimates particle sizes (due to the omission of relevant condensates, as shown in this work) and simulations are not evolved sufficiently in time to reach an equilibrium state of the precipitation (if one exists). Since the gravitational settling and radiative effects are so strongly linked to the particle sizes, using a cloud model that can explicitly solve for the condensate radii, such as CARMA \citep[see ][for its use in 1D]{powell18} could be very important. In future work, where appropriate comparisons can be made, we will present a more detailed comparison between simulations adopting both the microphysical and parameterised cloud models \cite[see ][for a comparison in 1D]{gao18a}.

As always, we are keen to highlight the key limitations of the simulations. We first note that atmospheric radiative adjustment may well depend on the initial conditions. For this work we start from a fully equilibrated cloud--free simulation of HD~209458b, but such an atmospheric state may not be reached in the presence of cloud from the first days of the planet's formation. One of the largest unknowns pertaining to the input physics is the choice of condensates. In this study we consider a select few species which are known to be important in equilibrium cloud models. However, there is a wide array of unknowns regarding the precise cloud species in the observable atmosphere; microphysically formed cloud could favour alternative cloud particle compositions for example, or some species may have settled out of the atmosphere completely having reached much larger physical radii than predicted here, and subsequently efficiently precipitated. We secondly caution the omission of cloud advection, which has been shown to be an important factor in the geometric distribution of cloud, as well as an influencing factor behind cloud--driven variability \citep{lines18a}. Finally, we acknowledge that the parameterisation of the vertical mixing in this work does not account for the vertical velocities arising from large scale circulation and may underestimate the abundance of cloud, particularly smaller particles, in the upper atmosphere. We endevour to address the influence of large scale flows on the cloud vertical distribution and particles sizes in future work.

\section*{Acknowledgements}
SL and JG are funded by and thankful to the Leverhulme Trust. NJM is part funded by a Leverhulme Trust Research Project Grant. JM and IAB acknowledge the support of a Met Office Academic Partnership secondment. This work was partly supported by a Science and Technology Facilities Council Consolidated Grant (ST/R000395/1). BD acknowledges support from a Science and Technology Facilities Council Consolidated Grant (ST/R000395/1). This research made use of the ISCA High Performance Computing Service at the University of Exeter. Material produced using Met Office Software. The authors are grateful to Mark Marley and Tiffany Kataria for access to the EddySed cloud formation code. SL extends thanks to T Lines for their programming insight. This work benefited from the 2018 Exoplanet Summer Program in the Other Worlds Laboratory (OWL) at the University of California, Santa Cruz, a program funded by the Heising--Simons Foundation. The research data supporting this publication are
openly available from the University of Exeter's institutional repository at: https://doi.org/10.24378/exe.1483 The authors are grateful for the excellent and insightful comments from the reviewer and an enjoyable review process.


\bibliographystyle{mnras}
\bibliography{clouds}

\end{document}